\documentclass[a4paper,12pt]{article}

\usepackage{amsmath,amssymb}
\usepackage{epsfig}
\usepackage{cite}

\makeatletter

\def\unit{{1\kern-.65ex {\rm l}}}
\def\1{{1\kern-.65ex {\rm l}}}
\def\ap{{\alpha'}}

\def\Im{\mathop{\mathrm{Im}}\nolimits}

\def\Re{\mathop{\mathrm{Re}}\nolimits}

\let\ev=\bracket
\let\Ev=\Bracket
\let\vev=\bracket

\def\Xt{{\widetilde{X}}}

\def\psit{{\widetilde{\psi}}}

\def\omegat{{\widetilde{\omega}}}

\def\CA{{\cal A}}

\def\CC{{\cal C}}
\def\CD{{\cal D}}

\def\CF{{\cal F}}

\def\CH{{\cal H}}

\def\CN{{\cal N}}
\def\CO{{\cal O}}

\def\CT{{\cal T}}

\def\bbR{{\mathbb{R}}}

\def\bbZ{{\mathbb{Z}}}

\begin{document}

\baselineskip=18pt  
\numberwithin{equation}{section}  



\thispagestyle{empty}

\vspace*{-2cm} 
\begin{flushright}
ITFA-2010-06
\end{flushright}

\vspace*{1cm} 
\begin{center}
 {\LARGE Holographic Brownian Motion and\\[1ex] Time Scales in Strongly Coupled Plasmas}\\
 \vspace*{1.7cm}
Ardian Nata Atmaja$^{1,2}$, Jan de Boer$^3$ and Masaki Shigemori$^{4,5}$\\
 \vspace*{1.0cm} 
 $^1$ 
Research Center for Physics, Indonesian Institute of Sciences (LIPI)\\
Kompleks PUSPITEK Serpong, Tangerang 15310, Indonesia\\[2ex]
$^2$ Indonesia Center for Theoretical and Mathematical Physics (ICTMP)\\
Bandung 40132, Indonesia\\[2ex]
 $^3$ Institute for Theoretical Physics, University of Amsterdam\\
 Valckenierstraat 65, 1018 XE Amsterdam, The Netherlands\\[2ex]
 $^4$ 
Yukawa Institute for Theoretical Physics (YITP), Kyoto University\\
Kitashirakawa Oiwakecho, Sakyo-ku, Kyoto 606-8502 Japan\\[2ex]
$^5$ Hakubi Center, Kyoto University\\
Yoshida-Ushinomiyacho, Sakyo-ku, Kyoto 606-8501, Japan\\
\end{center}
\vspace*{1cm}

\noindent
We study Brownian motion of a heavy quark in field theory plasma in the
AdS/CFT setup and discuss the time scales characterizing the interaction
between the Brownian particle and plasma constituents.  Based on a
simple kinetic theory, we first argue that the mean-free-path time is
related to the connected 4-point function of the random force felt by
the Brownian particle.  Then, by holographically computing the 4-point
function and regularizing the IR divergence appearing in the
computation, we write down a general formula for the mean-free-path
time, and apply it to the STU black hole which corresponds to plasma
charged under three $U(1)$ $R$-charges.  The result indicates that the
Brownian particle collides with many plasma constituents simultaneously.

\newpage
\setcounter{page}{1} 



\tableofcontents


\section{Introduction}
\label{sec:intro}

Brownian motion \cite{Brown, Uhlenbeck:1930zz, Duplantier} is a window
into the microscopic world of nature.  The random motion exhibited by a
small particle suspended on a fluid tells us that the fluid is not a
continuum but is actually made of constituents of finite size.  A
mathematical description of Brownian motion is given by the Langevin
equation, which phenomenologically describes the force acting on the
Brownian particle as a sum of dissipative and random forces.  Both of
these forces originate from the incessant collisions with the fluid
constituents and we can learn about the microscopic interaction between
the Brownian particle and the fluid constituents if we measure these
forces very precisely.  Brownian motion is a universal phenomenon in
finite temperature systems and any particle immersed in a fluid at
finite temperature undergoes Brownian motion; for example, a heavy quark
in the quark-gluon plasma also exhibits such motion.

The last several years have seen a considerable success in the
application of the AdS/CFT correspondence \cite{Maldacena:1997re,
Gubser:1998bc, Witten:1998qj} to the study of strongly coupled systems,
in particular the quark-gluon plasma.  The quark-gluon plasma of QCD is
believed to be qualitatively similar to the plasma of $\CN=4$ super
Yang--Mills theory, which is dual to string theory in an AdS black hole
spacetime.  The analysis of scattering amplitudes in the AdS black hole
background led to the universal viscosity bound \cite{Kovtun:2004de},
which played an important role in understanding the physics of the
elliptic flow observed at RHIC\@.  On the other hand, the study of the
physics of trailing strings in the AdS spacetime explained the
dissipative and diffusive physics of a quark moving through a field
theory plasma, such as the diffusion coefficient and transverse momentum
broadening \cite{Herzog:2006gh, Liu:2006ug, Gubser:2006bz,
Herzog:2006se, CasalderreySolana:2006rq, Gubser:2006nz, Liu:2006he,
CasalderreySolana:2007qw}.  The relation between the hydrodynamics of
the field theory plasma and the bulk black hole dynamics was first
revealed in \cite{Bhattacharyya:2008jc} (see also \cite{Son:2007vk}).

A quark immersed in a quark-gluon plasma exhibits Brownian motion.
Therefore, it is a natural next step to study Brownian motion using the
AdS/CFT correspondence.  An external quark immersed in a field theory
plasma corresponds to a bulk fundamental string stretching between the
boundary at infinity and the event horizon of the AdS black hole.  In
the finite temperature black hole background, the string undergoes a
random motion because of the Hawking radiation of the transverse
fluctuation modes \cite{Lawrence:1993sg, Rey:1998ik, Myers:2007we}.
This is the bulk dual of Brownian motion, as was clarified in
\cite{deBoer:2008gu, Son:2009vu}.  By studying the random motion of the
bulk ``Brownian string'', Refs.\ \cite{deBoer:2008gu, Son:2009vu}
derived the Langevin equation describing the random motion of the
external quark in the boundary field theory and determined the
parameters appearing in the Langevin equation.  Other recent work on
Brownian motion in AdS/CFT includes \cite{Giecold:2009cg,
Giecold:2009wi, CasalderreySolana:2009rm, Gursoy:2010aa}.

As mentioned above, by closely examining the random force felt by the
Brownian particle, we can learn about the interaction between the
Brownian particle and plasma constituents. The main purpose of the
current paper is to use the AdS/CFT dictionary to compute the
correlation functions of the random force felt by the boundary Brownian
particle by studying the bulk Brownian string.  From the random force
correlators, we can read off time scales characterizing the interaction
between the Brownian particle and plasma constituents, such as the
mean-free-path time $t_{\text{mfp}}$.  The computation of
$t_{\text{mfp}}$ has already been discussed in \cite{deBoer:2008gu} but
there it was partly based on dimensional analysis and the current paper
attempts to complete the computation.

More specifically, we will compute the 2- and 4-point functions of the
random force from the bulk and, based on a simple microscopic model,
relate them to the mean-free-path time $t_{\text{mfp}}$.  More
precisely, the time scale $t_{\text{mfp}}$ is related to the
non-Gaussianity of the random force statistics.  The computation of the
4-point function can be done using the standard
Gubser-Klebanov-Polyakov-Witten (GKPW) rule \cite{Gubser:1998bc,
Witten:1998qj} and holographic renormalization (as reviewed in {\it
e.g.}\ \cite{Skenderis:2002wp}) with the Lorentzian AdS/CFT prescription
of {\it e.g.}\ \cite{Skenderis:2008dg, vanRees:2009rw}. In the
computation, however, we encounter an IR divergence.  This is because we
are expanding the Nambu--Goto action in the transverse fluctuation
around a static configuration and the expansion breaks down very near
the horizon where the local temperature becomes of the string scale. We
regularize this IR divergence by cutting off the geometry near the
horizon at the point where the expansion breaks down.  For the case of a
neutral plasma, the resulting mean-free-path time is
\begin{align}
 t_{\text{mfp}}\sim {1\over T\log\lambda},\qquad
 \lambda\equiv{l^4\over\ap^2},
\end{align}
where $T$ is the temperature and $l$ is the AdS radius.  Because the
time elapsed in a single event of collision is $t_{\text{coll}}\sim
1/T$, this implies that the Brownian particle is undergoing
$\sim\log\lambda$ collisions simultaneously. (So, the term
mean-free-path time is probably a misnomer; it might be more appropriate
to call $t_{\text{mfp}}^{-1}$ the collision frequency instead.) We write
down a formula for $t_{\text{mfp}}$ for more general cases with
background charges. We apply it to the STU black hole which corresponds
to a plasma that carries three $U(1)$ $R$-charges.  This corresponds to
a situation where chemical potentials for baryon numbers have been
turned on.

\bigskip
The organization of the rest of the paper is as follows.  In section
\ref{sec:BM_in_AdS/CFT}, we start with a brief review of Brownian motion
in the AdS/CFT setup, from both the boundary and bulk viewpoints, taking
neutral AdS black holes as simple examples. Then we will discuss
Brownian motion in more general cases where the background plasma is
charged.  In section \ref{sec:time_scales}, we discuss various time
scales that characterize the interaction between the Brownian particle
and plasma constituents.  In particular, we introduce the mean-free-path
time $t_{\text{mfp}}$, which is the main objective of the current paper,
and relate it to the non-Gaussianity of the random force statistics
using a simple microscopic model.  In section \ref{sec:holoR4}, we use
the AdS/CFT correspondence to compute the random force correlators that
are necessary to obtain $t_{\text{mfp}}$.  We present two methods to
compute the correlation functions.  The first one is to treat the
worldsheet theory as a usual thermal field theory.  The second one is to
use the standard GKPW prescription and holographic renormalization
applied to the Lorentzian black hole backgrounds.  The expressions for
the random correlators turn out to be IR divergent.  In section
\ref{sec:IRdiv}, we discuss the physical meaning of this IR divergence
and propose a way to regularize it by cutting off the black hole
geometry near the horizon.  In section \ref{sec:gen's}, we write down
the formula for $t_{\text{mfp}}$ for general black holes and, as an
example, compute $t_{\text{mfp}}$ for a 3-charge black hole, the STU
black hole. Section \ref{sec:conclusions} is devoted to discussions.
The appendices contain details of the various computations in the main
text.

\section{Brownian motion in AdS/CFT}
\label{sec:BM_in_AdS/CFT}

In this section we will briefly review how Brownian motion is realized
in the AdS/CFT setup \cite{deBoer:2008gu, Son:2009vu}, mostly following
\cite{deBoer:2008gu}.  If we put an external quark in a CFT plasma at
finite temperature, the quark undergoes Brownian motion as it is kicked
around by the constituents of the plasma.  On the bulk side, this
external quark corresponds to a fundamental string stretching between
the boundary and the horizon.  This string exhibits random motion due to
Hawking radiation of its transverse modes, which is the dual of the
boundary Brownian motion.

We will explain the central ideas of Brownian motion in AdS/CFT using
the simple case where the background plasma is neutral.  In explicit
computations, we consider the AdS$_3$/CFT$_2$ example for which exact
results are available.  Then we will move on to discuss more general
cases of charged plasmas.

\subsection{Boundary  Brownian motion}
\label{subsec:bndy_BM}

Let us begin our discussion of Brownian motion from the boundary side,
where an external quark immersed in the CFT plasma undergoes random
Brownian motion.  A general formulation of non-relativistic Brownian
motion is based on the generalized Langevin equation \cite{Kubo:f-d_thm,
Mori:genLE}, which takes the following form in one spatial dimension:
\begin{equation}
 \dot p(t)=-\int_{-\infty}^t dt'\, \gamma(t-t')\, p(t')+R(t)+K(t),
 \label{genLE}
\end{equation}
where $p=m\dot x$ is the (non-relativistic) momentum of the Brownian
particle at position $x$, and $\dot{}\equiv d/dt$.  The first term on
the right hand side of (\ref{genLE}) represents (delayed) friction,
which depends linearly on the past trajectory of the particle via the
memory kernel $\gamma(t)$.  The second term corresponds to the random
force which we assume to have the following average:
\begin{equation}
 \vev{R(t)}=0,\qquad
 \vev{R(t)R(t')}=\kappa(t-t'),\label{RR=kappa}
\end{equation}
where $\kappa(t)$ is some function.  The random force is assumed to be
Gaussian; namely, all higher cumulants of $R$ vanish.  $K(t)$ is an
external force that can be added to the system.  The separation of the
force into frictional and random parts on the right hand side of
(\ref{genLE}) is merely a phenomenological simplification;
microscopically, the two forces have the same origin (collision with the
fluid constituents).  As a result of the two competing forces, the
Brownian particle exhibits thermal random motion.  The two functions
$\gamma(t)$ and $\kappa(t)$ completely characterize the Langevin
equation (\ref{genLE}).  Actually, $\gamma(t)$ and $\kappa(t)$ are
related to each other by the fluctuation-dissipation theorem \cite{KTH}.

The time evolution of the displacement squared of a Brownian particle
obeying (\ref{genLE}) has the following asymptotic behavior \cite{Uhlenbeck:1930zz}:
\begin{equation}
 \vev{s(t)^2}\equiv
 \vev{[x(t)-x(0)]^2}
 \approx
\begin{cases}
   \displaystyle {T\over m}t^2 &~~~  (t\ll t_{\rm relax}) \text{~~:~~ballistic regime}\\[3ex]
  \displaystyle 2Dt            &~~~  (t\gg t_{\rm relax}) \text{~~:~~diffusive regime}
\end{cases} 
 \label{s^2_simple}
\end{equation}
The crossover time scale $t_{\rm relax}$ between two regimes is given by
\begin{equation}
 {t_{\rm relax}}
  ={1\over \gamma_0},\qquad
  \gamma_0\equiv \int_0^\infty dt\,\gamma(t),
\label{t_relax_gen}
\end{equation}
while the diffusion constant $D$ is given by
\begin{equation}
 D={T\over\gamma_0 m}.
 \label{diff_const_def}
\end{equation}
In the ballistic regime, $t\ll t_{\rm relax}$, the particle moves
inertially ($s\sim t$) with the velocity determined by equipartition,
$|\dot x|\sim \sqrt{T/m}$, while in the diffusive regime, $t\gg t_{\rm
relax}$, the particle undergoes a random walk ($s\sim\sqrt{t}$).  This
is because the Brownian particle must be hit by a certain number of
fluid particles to lose the memory of its initial velocity.  The time
$t_{\rm relax}$ between the two regimes is called the relaxation time
which characterizes the time scale for the Brownian particle to
thermalize.

By Fourier transforming the Langevin equation \eqref{genLE}, we obtain
\begin{align}
 p(\omega)=\mu(\omega)[R(\omega)+K(\omega)],\qquad
 \mu(\omega)={1\over\gamma[\omega]-i\omega}.\label{admittance}
\end{align}
The quantity $\mu(\omega)$ is called the admittance which describes the
response of the Brownian particle to
perturbations. $p(\omega),R(\omega),K(\omega)$ are Fourier transforms,
{\it e.g.},
\begin{align}
 p(\omega)=\int_{-\infty}^\infty dt\, p(t)\,e^{i\omega t},
 \label{Fourier_trfm}
\end{align}
while
$\gamma[\omega]$ is the Fourier--Laplace transform:
\begin{align}
 \gamma[\omega]=\int_{0}^\infty dt\, \gamma(t) \,e^{i\omega t}.
 \label{FL_trfm}
\end{align}
In particular, if there is no external force, $K=0$, \eqref{admittance}
gives
\begin{align}
  p(\omega)=-im\omega  x(\omega)=\mu(\omega)R(\omega)
\end{align}
and, with the knowledge of $\mu$, we can determine the correlation functions
of the random
force $R$ from those of $p$ or those of the position $x$.

In the above, we discussed the Langevin equation in one spatial
dimension, but generalization to $n=d-2$ spatial dimensions is
straightforward.\footnote{We assume that $d\ge 3$ and thus $n\ge 1$.}

\subsection{Bulk Brownian motion}
\label{subsec:bulk_BM}

The AdS/CFT correspondence states that string theory in AdS$_{d}$ is
dual to a CFT in $(d-1)$ dimensions.  In particular, the neutral planar
AdS-Schwarzschild black hole with metric
\begin{align}
 ds_d^2&={r^2\over l^2}\left[-f(r)dt^2+(d X^I)^2\right]
 +{l^2\over r^2 f(r)}  dr^2,\qquad
 f(r)=1-\left({r_H\over r}\right)^{d-1}
 \label{AdSdBH}
\end{align}
is dual to a neutral CFT plasma at a temperature equal to the Hawking
temperature of the black hole,
\begin{align}
 T={1\over\beta}={(d-1)r_H\over 4\pi l^2}.
 \label{Hawk_temp_d}
\end{align}
In the above, $l$ is the AdS radius, $t\in\bbR$ is time, and
$X^I=(X^1,\dots, X^{d-2})\in \bbR^{d-2}$ are the spatial coordinates on
the boundary.  We will set $l=1$ henceforth.

\begin{figure}[tb]
\begin{center}
 \epsfig{file=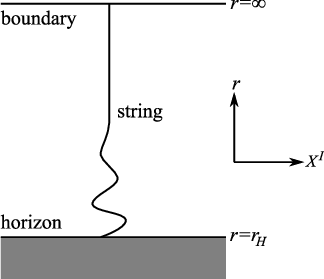,width=5cm} \caption{The bulk dual of a
Brownian particle: a fundamental string attached to the boundary of the
AdS space and dipping into the horizon. Because of the Hawking radiation
of the transverse fluctuation modes on the string, the string endpoint
at infinity moves randomly, corresponding to the Brownian motion on the
boundary. } \label{fig:brownianf1}
\end{center}
\end{figure}

The external quark in CFT corresponds in the bulk to a fundamental
string in the black hole geometry \eqref{AdSdBH} which is attached to the
boundary at $r=\infty$ and dips into the black hole horizon at
$r=r_H$; see Figure \ref{fig:brownianf1}.  The $X^I$
coordinates of the string at $r=\infty$ in the bulk define the boundary
position of the external quark.  As we discussed above, such an external
particle at finite temperature $T$  undergoes Brownian
motion. The bulk dual statement is that the black hole environment in
the bulk excites the modes on the string and, as the result, the
endpoint of the string at $r=\infty$ exhibits a Brownian motion which
can be modeled by a Langevin equation.

The string in the bulk does not just describe an external point-like
quark in the CFT with its position given by the position of the string
endpoint at $r=\infty$.  The transverse fluctuation modes of the bulk
string correspond on the CFT side to the degrees of freedom that were
induced by the injection of the external quark into the plasma.  In
other words, the quark immersed in the plasma is dressed with a
``cloud'' of excitations of the plasma and the transverse fluctuation
modes on the bulk string correspond to the excitation of this
cloud.\footnote{For recent discussions on this non-Abelian ``dressing'',
see \cite{Dominguez:2008vd, Beuf:2008ep, Chernicoff:2009ff}.}  In a
sense, the quark forms a ``bound state'' with the background plasma and
the excitation of the transverse fluctuation modes on the bulk string
corresponds to excited bound states.

We study this motion of a string in the probe approximation where we
ignore its backreaction on the background geometry.  We also assume that
there is no $B$-field in the background.  In the black hole geometry,
the transverse fluctuation modes of the string get excited due to
Hawking radiation \cite{Lawrence:1993sg}.  If the string coupling $g_s$ is small, we can ignore
the interaction between the transverse modes on the string and the
thermal gas of closed strings in the bulk of the AdS space.  This is
because the magnitude of Hawking radiation (for both string transverse
modes and the bulk closed strings) is controlled by $G_N\propto g_s^2$,
and the effect of the interaction between the transverse modes on the
string and the bulk modes is further down by $g_s^2$.

Let the string be along the $r$ direction and consider small fluctuations
of it in the transverse directions $X^I$.  The action for the string is
simply the Nambu--Goto action in the absence of a $B$-field.  In the gauge
where the world-sheet coordinates are identified with the spacetime
coordinates $x^\mu=t,r$, the transverse fluctuations $X^I$ become
functions of $x^\mu$: $X^I= X^I(x)$.  By expanding the Nambu--Goto action
up to quadratic order in $X^I$, we obtain
\begin{align}
 S_{\text{NG}}&=
 -{1\over2\pi\ap}\int d^2x\sqrt{-\det\gamma_{\mu\nu}}
 \approx
 {1\over 4\pi\ap}\int dt\,dr
 \biggl[
 {(\partial_t X^I)^2\over f}-{r^4f}\,(\partial_r X^I)^2
 \biggr]
 \equiv S_0,
\label{S_0}
\end{align}
where $\gamma_{\mu\nu}$ is the induced metric.  In the second
approximate equality we also dropped the constant term that does not
depend on $X^I$.  This quadratic approximation is valid as long as the
scalars $X^I$ do not fluctuate too far from their equilibrium value
(taken here to be $X^I =0$). This corresponds to taking a
non-relativistic limit for the transverse fluctuations.  We will be
concerned with the validity of this quadratic approximation later.  The
equation of motion derived from \eqref{S_0} is
\begin{align}
 [f^{-1}\omega^2+\partial_r(r^4f\partial_r )]X^I=0,
 \label{EOMgen}
\end{align}
where we set $X^I(r,t)\propto e^{-i\omega t}$.
Because $X^I$ with different polarizations $I$ are independent and
equivalent, we will consider only one of them, say $X^1$, and simply
call it $X$ henceforth.

The quadratic action \eqref{S_0} and the equation of motion
\eqref{EOMgen} derived from it are similar to those for a Klein--Gordon
scalar.  Therefore, the quantization of this theory can be done just the
same way, by expanding $X$ in a basis of solutions to
\eqref{EOMgen}. Because $t$ is an isometry direction of the geometry
\eqref{AdSdBH}, we can take the frequency $\omega$ to label the basis of
solutions.  So, let $\{u_{\omega}(x)\}$, $\omega>0$ be a basis of
positive-frequency modes.  Then we can expand $X$ as
\begin{align}
 X^I(x)=\int_0^\infty {d\omega\over 2\pi}
 [a_{\omega} u_{\omega}(x)+a_{\omega}^\dagger u_{\omega}(x)^*].\label{expnXI_gen}
\end{align}
If we normalize $u_{\omega}(x)$ by introducing an appropriate norm (see
Appendix \ref{sec:normalizing}), the operators $a,a^\dagger$ satisfy
the canonical commutation relation
\begin{align}
  [a_{\omega},a_{\omega'}]=
  [a_{\omega}^\dagger,a_{\omega' }^\dagger]=0,\qquad
  [a_{\omega},a_{\omega'}]=2\pi\delta(\omega-\omega').\label{CCR_a}
\end{align}

To determine the basis $\{u_{\omega}(x)\}$, we need to impose some
boundary condition at $r=\infty$.  The usual boundary condition in
Lorentzian AdS/CFT is to require normalizability of the modes at
$r=\infty$ \cite{Balasubramanian:1998sn} but, in the present case, that
would correspond to an infinitely long string extending to $r=\infty$,
which would mean that the mass of the external quark is infinite and
there would be no Brownian motion.  So, instead, we introduce a UV
cut-off\,\footnote{We use the terms ``UV'' and ``IR'' with respect to
the boundary energy.  In this terminology, in the bulk, UV means near
the boundary and IR means near the horizon.}  near the boundary to make
the mass very large but finite.  Specifically, we implement this by
means of a Neumann boundary condition
\begin{align}
\partial_r X=0\qquad \text{at}\quad r=r_c\gg r_H,\label{Neumanbc_rc}
\end{align}
where $r=r_c$ is the cut-off surface.\footnote{In the AdS/QCD context, one can
think of the cut-off being determined by the location of the flavour
brane, whose purpose again is to introduce dynamical (finite mass)
quarks into the field theory.}  The relation between this UV cut-off
$r=r_c$ and the mass $m$ of the external particle is easily computed
from the tension of the string:
\begin{align}
 m&= {1\over 2\pi\ap}\int_{r_H}^{r_c} dr\,\sqrt{g_{tt}\,g_{rr}}
 ={r_c-r_H\over 2\pi\ap}
 \approx{r_c\over 2\pi\ap}.
 \label{m_and_rc}
\end{align}

Before imposing a boundary condition, the wave equation \eqref{EOMgen}
in general has two solutions, which are related to each other by
$\omega\leftrightarrow -\omega$.  Denote these solutions by
$g_{\pm\omega}(r)$.  They are related by
$g_{\omega}(r)^*=g_{-\omega}(r)$.  These solutions are easy to obtain in
the near horizon region $r\approx r_H$, where the wave equation reduces
to
\begin{align}
 (\omega^2+\partial_{r_*}^2)X_\omega\approx 0.\label{eom_NH}
\end{align}
Here, $r_*$ is the tortoise coordinate defined by
\begin{align}
 dr_*={dr\over r^2f(r)}.\label{tortoise_gen}
\end{align}
Near the horizon, we have
\begin{align}
 r_*\sim {1\over (d-1)r_H}\log\left({r-r_H\over r_H}\right)
 \label{tortoise_NH}
\end{align}
up to an additive numerical constant. Normally this constant is fixed by
setting $r_*=0$ at $r=\infty$, but we will later find that some other
choice is more convenient.  From \eqref{eom_NH}, we see that, in the
near horizon region $r=r_H$, we have the following outgoing and ingoing
solutions:
\begin{align}
 g_\omega(r)\approx e^{i\omega r_*}:~\text{outgoing},\qquad
 g_{-\omega}(r)\approx e^{-i\omega r_*}:~\text{ingoing}.\label{g_NH_asympt}
\end{align}
The boundary condition \eqref{Neumanbc_rc} dictates that we take the
linear combination
\begin{align}
 f_\omega(r)
 =g_{\omega}(r)
 +e^{i\theta_\omega}g_{-\omega}(r),\qquad
 e^{i\theta_\omega}
 =-{\partial_r g_{\omega}(r_c) \over \partial_r g_{-\omega}(r_c)}.
 \label{f_ito_g}
\end{align}
We can show that $\theta_\omega$ is real using the fact that
$g_{-\omega}=g_\omega^*$.

The normalized modes $u_\omega(t,r)$ are essentially given by
$f_\omega(r)$; namely, $u_\omega(t,r)\propto e^{-i\omega t}f_\omega(r)$.
A short analysis of the norm (see Appendix \ref{sec:normalizing}) shows
that the correctly normalized mode expansion is given by
\begin{align}
 X(t,r)={\sqrt{2\pi\ap}\over r_H}\int_0^\infty {d\omega\over 2\pi}
 {1\over \sqrt{2\omega}}
 \left[ f_\omega(r)e^{-i\omega t}a_\omega+ f_\omega(r)^* e^{i\omega t}a_\omega^\dagger\right],\label{Xexpn_neutral}
\end{align}
where $f_\omega(r)$ behaves near the horizon as
\begin{align}
 f_\omega(r)\to e^{i\omega r_*} +e^{i\theta_\omega}e^{-i\omega r_*},\qquad
 r\to r_H\quad (r_*\to-\infty).\label{f_NH_asympt}
\end{align}
If we can find such $f_\omega(r)$, then $a,a^\dagger$ satisfy the
canonically normalized commutation relation \eqref{CCR_a}.

We identify the position $x(t)$ of the boundary Brownian particle with
$X(t,r)$ at the cutoff $r=r_c$:
\begin{align}
 x(t)\equiv X(t,r_c)={\sqrt{2\pi\ap}\over r_H}\int_0^\infty {d\omega\over 2\pi}
 {1\over \sqrt{2\omega}}
 [ f_\omega(r_c)e^{-i\omega t}a_\omega+ f_\omega(r_c)^* e^{i\omega t}a_\omega^\dagger].
 \label{rel_x_aadagger}
\end{align}
The equation \eqref{rel_x_aadagger} relates the correlation functions of
$x(t)$ to those of $a,a^\dagger$.  Because the quantum field $X(t,r)$ is
immersed in a black hole background, its modes Hawking radiate
\cite{Lawrence:1993sg}. This can be seen from the fact that, near the
horizon, the worldsheet action \eqref{S_0} is the same as that of a
Klein--Gordon field near a two-dimensional black hole. The standard
quantization of fields in curved spacetime \cite{Birrell:1982ix} shows
that the field gets excited at the Hawking temperature.  At the
semiclassical level, the excitation is purely thermal:
\begin{align}
 \ev{a^\dagger_\omega a_{\omega'}}={2\pi\delta(\omega-\omega')\over e^{\beta\omega}-1}.\label{aad_therm_exp}
\end{align}
Using \eqref{rel_x_aadagger} and \eqref{aad_therm_exp}, one can compute the
correlators of $x$ to show that it undergoes Brownian motion
\cite{deBoer:2008gu}, having both the ballistic and diffusive regimes.

In the AdS$_3$ ($d=3$) case, we can carry out the above procedure very
explicitly.  In this case, the metric \eqref{AdSdBH} becomes the
nonrotating BTZ black hole:
\begin{align}
 ds^2=
 -{(r^2-r_H^2)}\, dt^2
 +{dr^2\over r^2-r_H^2}
 +{r^2}\, dX^2.
 \label{nonrotBTZ}
\end{align}
For the usual BTZ black hole, $X$ is written as $X=\phi$ where
$\phi\cong\phi+2\pi$, but here we are taking $X\in\bbR$, corresponding
to a ``planar'' black hole.  The Hawking temperature \eqref{Hawk_temp_d}
is, in this case,
\begin{align}
 T\equiv{1\over\beta} ={r_H\over 2\pi}.
 \label{Hawk_temp}
\end{align}
In terms of the tortoise coordinate $r_*$,
the metric \eqref{nonrotBTZ} becomes
\begin{align}
 ds^2=(r^2-r_H^2) (-dt^2+dr_*^2)+{r^2}\, dX^2,
 \quad\qquad
 r_*\equiv{1\over 2r_H}\ln\left({r-r_H\over r+r_H}\right).
\label{tortoise}
\end{align}
The linearly independent solutions to \eqref{EOMgen} are given by
$g_{\pm\omega}(r)$, where
\begin{gather}
 g_{\omega}(r)
 = {1\over 1+ i\nu}{\rho+ i\nu \over \rho}\left({\rho-1\over \rho+1}\right)^{i\nu/2}
 = {1\over 1+ i\nu}{\rho+ i\nu \over \rho}\, e^{i\omega r_*}.
\label{fpm_def}
\end{gather}
Here we introduced
\begin{align}
 \rho\equiv {r\over r_H},\qquad
 \nu\equiv{\omega\over r_H}={\beta \omega\over 2\pi}.
 \label{nu_def}
\end{align}
The linear combination that satisfies the Neumann boundary condition
\eqref{Neumanbc_rc} is
\begin{align}
\begin{split}
  f_\omega &=
 g_\omega(\rho) + e^{i\theta_\omega} g_{-\omega}(\rho),\\
 e^{i\theta_\omega}
 &=-{\partial_r g_{\omega}(r_c)\over \partial_r g_{-\omega}(r_c)}
 = {1-i\nu\over 1+i\nu }\,{1 + i \rho_c\nu\over 1-i \rho_c\nu }
 \left({\rho_c-1\over \rho_c+1}\right)^{i\nu},
\end{split} 
\label{BoverA}
\end{align}
where $\rho_c\equiv r_c/r_H$.  This has the correct near-horizon
behavior \eqref{f_NH_asympt} too.  

By analyzing the correlators of $x(t)$ using the bulk Brownian motion,
one can determine the admittance $\mu(\omega)$ defined in
\eqref{admittance} for the dual boundary Brownian motion
\cite{deBoer:2008gu}. Although the result for general frequency $\omega$
is difficult to obtain analytically for general dimensions $d$, its
low-frequency behavior is relatively easy to find; this was done in
\cite{deBoer:2008gu} and the result for AdS$_d$/CFT$_{d-1}$ is
\begin{align}
 \mu(\omega)&={(d-1)^2 \ap\beta^2m\over 8\pi}+\CO(\omega).\label{mu_gen}
\end{align}
This agrees with the results obtained by drag force computations
\cite{Herzog:2006gh, Liu:2006ug, Gubser:2006bz,
CasalderreySolana:2006rq}.  For later use, let us also record the
low-frequency behavior of the random force correlator obtained in
\cite{deBoer:2008gu}:
\begin{align}
 G^{(R)}(t_1,t_2)&\equiv\ev{\CT[R(t_1)R(t_2)]},\\
 G^{(R)}(\omega_1,\omega_2)&= 2\pi\delta(\omega_1+\omega_2)
 \left[{16\pi\over (d-1)^2\ap \beta^3}+\CO(\omega)\right],
 \label{RRcorr}
\end{align}
where $T$ is the time ordering operator.

\subsection{Generalizations}
\label{subsec:BM_bulk_gen's}

In the above, we considered the simple case of neutral black holes,
corresponding to neutral plasmas in field theory.  More generally,
however, we can consider situations where the field theory plasmas carry
nonvanishing conserved charges.  For example, the quark-gluon plasma
experimentally produced by heavy ion collision has net baryon number.
Field theory plasmas charged under such global $U(1)$ symmetries
correspond on the AdS side to black holes charged under $U(1)$ gauge
fields.

On the gravity side of the correspondence, we do not just have AdS$_d$
space but also some internal manifold on which higher-dimensional
string/M theory has been compactified.  $U(1)$ gauge fields in the
AdS$_d$ space can be coming from (i) form fields in higher dimensions
upon compactification on the internal manifold, or (ii) the off-diagonal
components of the higher dimensional metric with one index along the
internal manifold.
In the former case (i), a charged CFT plasma corresponds to a charged
black hole, {\it i.e.}\ a Reissner--Nordstr\"om black hole (or a
generalization thereof to form fields) in the full spacetime.  In this
case, the analysis in the previous subsections applies almost
unmodified, because a fundamental string is not charged under such form
fields (except for the $B$-field which is assumed to vanish in the
present paper) and its motion is not affected by the existence of those
form fields.  Namely, the same configuration of a string---stretching
straight between the AdS boundary and the horizon and trivial in the
internal directions---is a solution of the Nambu--Goto action.
Therefore, as far as the fluctuation in the AdS$_d$ directions is
concerned, we can forget about the internal directions and the analysis
in the previous subsections goes through unaltered, except that the
metric \eqref{AdSdBH} must be replaced by an appropriate AdS black hole
metric deformed by the existence of charges.

The latter case (ii), on the other hand, corresponds to having a
rotating black hole (Kerr black hole) in the full spacetime.  A notable
example is the STU black hole which is a non-rotating black hole
solution of five-dimensional AdS supergravity charged under three
different $U(1)$ gauge fields \cite{Behrndt:1998jd}.  From the point of
view of 10-dimensional Type IIB string theory in $AdS_5\times S^5$, this
black hole is a Kerr black hole with three angular momenta in the $S^5$
directions \cite{Cvetic:1999xp}.  This solution can also be obtained by
taking the decoupling limit of the spinning D3-brane metric
\cite{Kraus:1998hv, Russo:1998by, Cvetic:1999xp}.
Analyzing the motion of a fundamental string in such a background
spacetime in general requires a 10-dimensional treatment, because the
string gets affected by the angular momentum of the black hole in the
internal directions \cite{Herzog:2006se, Caceres:2006dj, Herzog:2007kh}.
So, to study the bulk Brownian motion in such situations, we have to
find a background solution in the full 10-dimensional spacetime and
consider fluctuation around that 10-dimensional configuration.  The
background solution is straight in the AdS part as before but can be
nontrivial in the internal directions due to the drag by the geometry.

In either case, to study the transverse fluctuation of the string around
a background configuration, we do not need the full 10- or
11-dimensional metric.  For simplicity, let us focus on the transverse
fluctuation in one of the AdS$_d$ directions.  Then we only need the
three-dimensional line element along the directions of the background
string configuration and the direction of the fluctuation.  Let us write
the three-dimensional line element in general as
\begin{equation}
\label{gen non-rotating metric}
 ds^2=-h_t(r) f(r) dt^2+\frac{h_r(r)}{f(r)}dr^2+G(r) dX^2.
\end{equation}
$X$ is one of the spatial directions in AdS$_d$, parallel to the
boundary.  It is assumed that $X(t,r)=0$ is a solution to the
Nambu--Goto action in the full (10- or 11-dimensional) spacetime, and we
are interested in the fluctuations around it.\footnote{Note that, under
this assumption in a static spacetime, the three-dimensional line
element can be always written in the form of \eqref{gen non-rotating
metric}.  The $(t,r)$ and $(t,X)$ components should vanish by the
assumption that $X(t,r)=0$ is a solution, and the $(t,r)$ component can
be eliminated by a coordinate transformation.} The nontrivial effects in
the internal directions have been incorporated in this metric \eqref{gen
non-rotating metric}.  We will see how such a line element arises in the
explicit example of the STU black hole in section \ref{sec:gen's}.  In
this subsection, we will briefly discuss the random motion of a string
in general backgrounds using the metric \eqref{gen non-rotating metric}.

In the metric \eqref{gen non-rotating metric}, the horizon is at $r=r_H$
where $r_H$ is the largest positive solution to $f(r)=0$. The functions
$h_t(r)$ and $h_r(r)$ are assumed to be regular and positive in the
range $r_H\leq r<\infty$. Near the horizon $r\approx r_H$, expand $f(r)$
as
\begin{equation}
 f(r)\approx 2k_H(r-r_H),~~~~~k_H\equiv\frac{1}{2}f'(r_H).
\end{equation}
The Hawking temperature of the black hole, $T$, is given by
\begin{equation}
 T={1\over\beta}
  =\frac{k_H}{2\pi}\sqrt{\frac{h_t(r_H)}{h_r(r_H)}}.
\label{Gen temp}
\end{equation}
For the metric to asymptote to AdS near the boundary, we have
\begin{equation}
 h_t f\sim\frac{r^2}{l^2},~~~~~\frac{h_r}{f}\sim\frac{l^2}{r^2}
  \qquad\text{as} \qquad r\to \infty , \label{asymptcond_h}
\end{equation}
where we reinstated the AdS radius $l$.  Also, because the $X$ direction
\eqref{gen non-rotating metric} is assumed to be one of the spatial
directions of the AdS$_d$ directions parallel to the boundary, $G(r)$
must go as
\begin{equation}
 G\sim\frac{r^2}{l^2}
  \qquad\text{as} \qquad r\to \infty.\label{asymptcond_G}
\end{equation}
We demand that $G(r)$ be regular and positive in the region
$r_H\leq r<\infty$.  Note that the parametrization of the two metric
components $g_{tt},g_{rr}$ using three functions $h_t,h_r,f$ is
redundant and thus has some arbitrariness.

Consider fluctuation around the background configuration $X(t,r)=0$ in
the static gauge where $t,r$ are the worldsheet coordinates.  Just as in
\eqref{S_0}, the quadratic action obtained by expanding the Nambu--Goto
action in $X$ is
\begin{align}
 S_0&=-\frac{1}{4\pi\alpha'}\int d\sigma^2\sqrt{-g}\,G\,g^{\mu\nu}\partial_\mu X\partial_\nu X,\label{Gen NG expansion}
\end{align}
where $g_{\mu\nu}$ is the $t,r$ part of the metric \eqref{gen
non-rotating metric} ({\it i.e.}, the induced worldsheet metric for the
background configuration $X(t,r)=0$), and $g=\det g_{\mu\nu}$.  The
equation of motion derived from the quadratic action \eqref{Gen NG
expansion} is
\begin{equation}
\label{gen eqm}
 -\ddot{X}+\sqrt{\frac{h_t}{h_r}}\,\frac{f}{G}\,
 \partial_r\!\left(\sqrt{\frac{h_t}{h_r}}\,f G X'\right)=0,
\end{equation}
where $\,\dot{}=\partial_t$, $'=\partial_r$.  In terms of the tortoise
coordinate $r_*$ defined by
\begin{align}
 dr_*=\frac{1}{f}\sqrt{\frac{h_r}{h_t}}\,dr,
 \label{tortoise_def_gen}
\end{align}
\eqref{gen eqm} becomes a Schrodinger-like wave equation
\cite{Harmark:2007jy}:
\begin{equation}
\label{gen equation}
 \left[\frac{d^2}{dr_*^2}+\omega^2-V(r)\right]X_\omega(r)=0,
\end{equation}
where we set $X(t,r)=e^{-i\omega t}\eta(r)X_\omega(r)$ and the
``potential'' $V(r)$ is given by
\begin{equation}
 V(r)=-\eta\frac{dr}{dr_*}\frac{d}{dr}\left[\frac{1}{\eta^2}\frac{dr}{dr_*}\frac{d\eta}{dr}\right],\qquad \eta=G^{-1/2}.
\end{equation}
The potential $V(r)$ vanishes at the horizon and will become more and
more important as we move towards the boundary $r\to\infty$ where
$V(r)\sim 2r^2/l^4$.

Just as in the previous subsection, let the two solutions to the wave
equation \eqref{gen equation} be $g_{\omega}(r)$ and
$g_{-\omega}(r)=g_\omega(r)^*$.  Near the horizon where $V(r)=0$, the
wave equation \eqref{gen equation} takes the same form as \eqref{eom_NH}
and therefore $g_{\pm\omega}(r)$ can be taken to have the following
behavior near the horizon
\begin{align}
 g_{\pm \omega}(r)\to e^{\pm i\omega r_*}
 \qquad\text{as}\quad r\to r_H.
 \label{g_NH_asympt_gen}
\end{align}
If we introduce a UV cutoff at $r=r_c$ as before, the solution
$f_\omega(r)$ satisfying the Neumann boundary condition
(\ref{Neumanbc_rc}) at $r=r_c$ is a linear combination of
$g_{\pm\omega}(r)$ and can be written as \eqref{f_ito_g}.  Using this
$f_\omega(r)$, we can expand the fluctuation field $X(t,r)$ as
\begin{align}
 X(t,r)=\sqrt{2\pi\ap \over G(r_H)}\int_0^\infty {d\omega\over 2\pi}
 {1\over \sqrt{2\omega}}
 \left[ f_\omega(r)e^{-i\omega t}a_\omega+ f_\omega(r)^* e^{i\omega t}a_\omega^\dagger\right],\label{Xexpn_gen}
\end{align}
where $a_\omega,a_\omega^\dagger$ are canonically normalized to satisfy
\eqref{CCR_a}.  As before, the value of $X(t,r)$ at the UV cutoff
$r=r_c$ is interpreted as the position $x(t)$ of the boundary Brownian
motion: $X(t,r_c)\equiv x(t)$.  By assuming that the modes Hawking
radiate thermally as in \eqref{aad_therm_exp}, we can determine the
parameters of the boundary Brownian motion such as the admittance
$\mu(\omega)$.

In general, solving the wave equation \eqref{gen equation} and obtaining
explicit analytic expressions for $g_{\pm\omega},f_\omega$ is difficult.
However, in the low frequency limit $\omega\to 0$, it is possible to
determine their explicit forms as explained in \cite{deBoer:2008gu} or
in Appendix \ref{sec:lowE_wave} and, based on that, one can compute the
low frequency limit of $\mu(\omega)$ following the procedure explained
in \cite{deBoer:2008gu}.  The result is
\begin{equation}
\mu(\omega)=\frac{2m\pi\alpha'}{G(r_H)}+\CO(\omega).\label{mu_gen_geo}
\end{equation}
From this, we can derive the low frequency limit of the random force
correlator as follows:
\begin{align}
 G^{(R)}(\omega_1,\omega_2)&= 2\pi\delta(\omega_1+\omega_2)
 \left[{G(r_H)\over \pi\ap \beta}+\CO(\omega)\right].
 \label{gen RR correlator}
\end{align}

\section{Time scales}
\label{sec:time_scales}

\subsection{Physics of time scales}

In Eq.\ \eqref{t_relax_gen}, we introduced the relaxation time
$t_{\text{relax}}$ which characterizes the thermalization time of the
Brownian particle.  From Brownian motion, we can read off other physical
time scales characterizing the interaction between the Brownian particle
and plasma.

One such time scale, the microscopic (or collision duration) time
$t_{\rm {coll}}$, is defined to be the width of the random force
correlator function $\kappa(t)$.  Specifically, let us define
\begin{eqnarray}
 t_{\rm {coll}} = \int_0^{\infty}\!\! dt \,{\kappa(t)\over \kappa(0)}.
 \label{def_t_coll}
\end{eqnarray}
If $\kappa(t)=\kappa(0)e^{-t/t_{\rm {coll}}}$, the right hand side of
this precisely gives $t_{\rm {coll}}$.  This $t_{\rm {coll}}$
characterizes the time scale over which the random force is correlated,
and thus can be interpreted as the time elapsed in a single process of
scattering.  In usual situations,
\begin{eqnarray}
 t_{\rm {relax}}\gg t_{\rm {coll}}.
\label{tr>>tc}
\end{eqnarray}

\begin{figure}[htb]
\begin{center}
 \epsfig{file=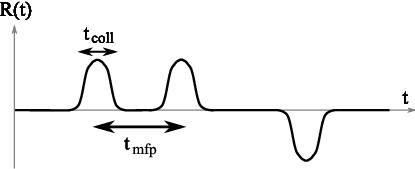,width=8cm} 
\caption{
 A sample of the stochastic variable $R(t)$, which consists of many pulses randomly distributed.}
 \label{fig:pulses}
\end{center}
\end{figure}

Another natural time scale is the mean-free-path time $t_{\rm {mfp}}$
given by the typical time elapsed between two collisions.  In the usual
kinetic theory, this mean free path time is typically $t_{\rm {coll}}
\ll t_{\rm {mfp}} \ll t_{\rm {relax}}$; however in the case of present
interest, this separation no longer holds, as we will see.  For a
schematic explanation of the timescales $t_{\rm coll}$ and $t_{\rm
mfp}$, see Figure \ref{fig:pulses}.

\subsection{A simple model}
\label{subsec:simple_model}

The collision duration time $t_{\text{coll}}$ can be read off from the
random force 2-point function $\kappa(t)=\ev{R(t)R(0)}$.  To determine
the mean-free-path time $t_{\text{mfp}}$, we need higher point functions
and some microscopic model which relates those higher point functions
with $t_{\text{mfp}}$.  Here we propose a simple model\,\footnote{This
is a generalization of the discussion given in Appendix D.1 of
\cite{deBoer:2008gu}. For somewhat similar models (binary collision
models), see \cite{Dunkel:2008} and references therein.} which relates
$t_{\text{mfp}}$ with certain 4-point functions of the random force $R$.

For simplicity, we first consider the case with one spatial dimension.
Consider a stochastic quantity $R(t)$ whose functional form consists of
many pulses randomly distributed.  $R(t)$ is assumed to be a classical
quantity (c-number).  Let the form of a single pulse be $P(t)$.
Furthermore, assume that the pulses come with random signs.  If we have
$k$ pulses at $t=t_i$ ($i=1,2,\dots,k$), then $R(t)$ is given by
\begin{align}
 R(t)&=\sum_{i=1}^k \epsilon_i P(t-t_i),
\label{R(T)sumofP}
\end{align}
where $\epsilon_i=\pm 1$ are random signs.

Let the distribution of pulses obey the Poisson distribution, which is a
physically reasonable assumption if $R$ is caused by random collisions.
This means that the probability that there are $k$ pulses in an interval
of length $\tau$, say $[0,\tau]$, is given by
\begin{align}
 P_k(\tau)=e^{-\mu\tau}{(\mu\tau)^k\over k!}.
\end{align}
Here, $\mu$ is the number of pulses per unit time.  In other words,
$1/\mu$ is the average distance between two pulses.  We do not assume
that the pulses are well separated; namely, we do \emph{not} assume
$\Delta\ll 1/\mu$.  If we identify $R(t)$ with the random force in the
Langevin equation, $t_{\text{mfp}}=1/\mu$.

The 2-point function for $R$ can be written as
\begin{align}
 \ev{R(t)R(t')}=\sum_{k=1}^\infty 
 e^{-\mu\tau}{(\mu\tau)^k\over k!}
 \sum_{i,j=1}^k \ev{\epsilon_i \epsilon_j P(t-t_i)P(t'-t_j)}_k,\label{bjse22Dec08}
\end{align}
where we assumed $t,t'\in[0,\tau]$ and $\ev{~}_k$ is the statistical
average when there are $k$ pulses in the interval $[0,\tau]$.  Because
$k$ pulses are randomly and independently distributed in the interval
$[0,\tau]$ by assumption, this expectation value is computed as
\begin{multline}
 \sum_{i,j=1}^k \ev{\epsilon_i \epsilon_j P(t-t_i)P(t'-t_j)}_k
\\
 =
 {1\over \tau^k}
 \int_0^\tau dt_1\cdots dt_k
 \left[
 \sum_{i=1}^k P(t-t_i)P(t'-t_i)
 +\sum_{i\neq j}^k \ev{\epsilon_i \epsilon_j}_k P(t-t_i)P(t'-t_j)
 \right].
\end{multline}
Here, the second term vanishes because $\ev{\epsilon_i\epsilon_j}_k=0$
for $i\neq j$.  Therefore, one readily computes
\begin{align}
 \sum_{i,j=1}\ev{\epsilon_i \epsilon_j P(t-t_i)P(t'-t_j)}_k
 &=
 {k\over \tau}
 \int_0^\tau dt_1 P(t-t_1)P(t'-t_1)\notag\\
 &\approx
 {k\over \tau}
 \int_{-\infty}^\infty  dt_1\, P(t-t_1)P(t'-t_1).
\end{align}
Here, in going to the second line, we took $\tau$ to be much larger than
the support of $P(t)$, which is always possible because $\tau$
is arbitrary. Substituting this back into \eqref{bjse22Dec08}, we find
\begin{align}
 \ev{R(t)R(t')}=\mu  \int_{-\infty}^\infty  dt_1\, P(t-t_1)P(t'-t_1).
\end{align}

In a similar way, one can compute the following 4-point function:
\begin{align}
 &\ev{R(t)R(t')R(t'')R(t''')}\notag\\
 &=\sum_{k=1}^\infty 
 e^{-\mu\tau}{(\mu\tau)^k\over k!}
 \sum_{i,j,m,n=1}^k 
 \ev{\epsilon_i \epsilon_j \epsilon_m \epsilon_n
 P(t-t_i)P(t'-t_j)P(t''-t_m)P(t'''-t_n)}_k.\label{blqk22Dec08}
\end{align}
Again, the expectation value $\ev{\epsilon_i \epsilon_j \epsilon_m
\epsilon_n}_k$ vanishes unless some of $i,j,m,n$ are equal.  The
possibilities are $i=j\neq m=n$, $i=m\neq j=n$, $i=n\neq j=m$, and
$i=j=m=n$.  Taking into account all these possibilities, in the end we
have
\begin{align}
 \ev{R(t)R(t')R(t'')R(t''')}
 &=
 \ev{R(t)R(t')R(t'')R(t''')}_{\text{disc}}
 +\ev{R(t)R(t')R(t'')R(t''')}_{\text{conn}},\label{RtRtRtRt}
\end{align}
where
\begin{align}
 \ev{R(t)R(t')R(t'')R(t''')}_{\text{disc}}
 &=
 \ev{R(t)R(t')}\ev{R(t'')R(t''')}+\ev{R(t)R(t'')}\ev{R(t')R(t''')}\notag\\
 &\qquad
 +\ev{R(t)R(t''')}\ev{R(t')R(t'')},\label{flay3Oct09}
 \\
 \ev{R(t)R(t')R(t'')R(t''')}_{\text{conn}}
 &=
 \mu \int_{-\infty}^\infty du \, P(t-u) P(t'-u) P(t''-u) P(t'''-u).
 \label{jtet8Apr09}
\end{align}
We can think of \eqref{flay3Oct09} as the ``disconnected part'' and
\eqref{jtet8Apr09} as the ``connected part'', or non-Gaussianity of
the random force statistics.

In the Fourier space, the expressions for these correlation functions
simplify:
\begin{align}
 \ev{R(\omega_1)R(\omega_2)}&
 =2\pi \mu \delta(\omega_1+\omega_2)P(\omega_1)P(\omega_2),
  \label{RR1dim_freq}
 \\
 \ev{R(\omega_1)R(\omega_2)R(\omega_3)R(\omega_4)}_{\text{disc}}
 &=
 (2\pi\mu)^2[
 \delta(\omega_1+\omega_2)\delta(\omega_3+\omega_4)
 +\delta(\omega_1+\omega_3)\delta(\omega_2+\omega_4)\notag\\
 &\qquad\qquad
 +\delta(\omega_1+\omega_4)\delta(\omega_2+\omega_3)
]
  P(\omega_1)P(\omega_2)P(\omega_3)P(\omega_4),
 \notag\\
 \ev{R(\omega_1)R(\omega_2)R(\omega_3)R(\omega_4)}_{\text{conn}}
 &=
 2\pi\mu \delta(\omega_1+\omega_2+\omega_3+\omega_4)
  P(\omega_1)P(\omega_2)P(\omega_3)P(\omega_4).
 \label{RRRR1dim_freq}
\end{align}
In particular, for small $\omega_i$,
\begin{align}
 \ev{R(\omega_1)R(\omega_2)}
 &\approx 2\pi \mu \delta(\omega_1+\omega_2)P(\omega=0)^2\label{jxcu15Sep09}\\
 \ev{R(\omega_1)R(\omega_2)R(\omega_3)R(\omega_4)}_{\text{conn}}
 &\approx
 2\pi\mu \delta(\omega_1+\omega_2+\omega_3+\omega_4)
 P(\omega=0)^4.\label{jvmx8Apr09}
\end{align}
Therefore, from the small frequency behavior of 2-point function and
connected 4-point function, we can separately read off the
mean-free-path time $t_{\text{mfp}}\sim 1/\mu$ and $P(\omega=0)$, the
impact per collision.

The discussion thus far has been focused on the case with one spatial
dimension, but generalization to $n=d-2$ spatial dimensions is
straightforward.  In this case, the random force becomes an
$n$-dimensional vector $R^I(t)$, $I=1,2,\dots,n$.  Generalizing
\eqref{R(T)sumofP}, let us model the random force to be given by a sum
of pulses:
\begin{align}
 R^I(t)=\sum_{i=1}^k \epsilon^I_i P(t-t_i).
\end{align}
Here, for each value of $i$, $\epsilon^I_i$ is a stochastic variable
taking random values in the $(n-1)$-dimensional sphere $S^{n-1}$.  We
also assume that $\epsilon^I_i$ for different values of $i$ are
independent of each other.  Then we can readily compute the following
statistical average:
\begin{align}
 \ev{\epsilon_i^I\epsilon_i^J}={\delta^{IJ}\over n},\qquad
 \ev{\epsilon_i^I \epsilon_i^J \epsilon_i^K \epsilon_i^L}=
 {\delta^{IJ}\delta^{KL}+\delta^{IK}\delta^{JL}+\delta^{IL}\delta^{JK}\over n(n+2)}.\label{ev_ee,eeee}
\end{align}
From this, we can derive the following $R$-correlators:
\begin{align}
 \ev{R^I(\omega_1)R^J(\omega_2)}&
 ={2\pi \mu \over n}\delta^{IJ}\delta(\omega_1+\omega_2)P(\omega_1)P(\omega_2),\\
 \ev{R^I(\omega_1)R^J(\omega_2)R^K(\omega_3)R^L(\omega_4)}
 &=
 \ev{R^I(\omega_1)R^J(\omega_2)R^K(\omega_3)R^L(\omega_4)}_{\text{conn}}\notag\\
 &\qquad+
 \ev{R^I(\omega_1)R^J(\omega_2)R^K(\omega_3)R^L(\omega_4)}_{\text{disc}},
\end{align}
where
\begin{align}
 \ev{R^I(\omega_1)R^J(\omega_2)R^K(\omega_3)R^L(\omega_4)}_{\text{disc}}
 &=
 \ev{R^I(\omega_1)R^J(\omega_2)}\ev{R^K(\omega_3)R^L(\omega_4)}\notag\\
 &\quad
 +\ev{R^I(\omega_1)R^K(\omega_3)}\ev{R^J(\omega_2)R^L(\omega_4)}\notag\\
 &\quad
 +\ev{R^I(\omega_1)R^L(\omega_4)}\ev{R^J(\omega_2)R^K(\omega_3)},
 \\
 \ev{R^I(\omega_1)R^J(\omega_2)R^K(\omega_3)R^L(\omega_4)}_{\text{conn}}
 &
 =
 {2\pi\mu\over n(n+2)}
 (\delta^{IJ}\delta^{KL}+\delta^{IK}\delta^{JL}+\delta^{IL}\delta^{JK})\notag\\
 &\quad\times
 \delta(\omega_1+\omega_2+\omega_3+\omega_4)
  P(\omega_1)P(\omega_2)P(\omega_3)P(\omega_4).\label{RIRJRKRLconn}
\end{align}
These are essentially the same as the $n=1$ results \eqref{RR1dim_freq},
\eqref{RRRR1dim_freq} and we can compute the mean-free-path time
$t_{\text{mfp}}\sim 1/\mu$ from the small $\omega$ behavior of 2- and
4-point functions.


One may wonder about the validity of the simple classical model we
proposed here, because of the various simplifications and assumptions
we made.  For example, we assumed that the distribution of pulses
is given by the Poisson distribution.  This is a natural assumption but,
in real systems, different pulses might be correlated and the deviation
of the distribution from the Poisson distribution may be appreciable.
Also, our model is classical whereas in the real system quantum effects
may not be ignorable.  In the first place, the kinetic theory picture of
independent particles colliding with each other is based on weak
coupling intuition and in strongly coupled systems its validity is
unclear.
However, the simplicity of our model can be regarded as its strength
too.  Because of its simplicity, our model can be thought of as a zeroth
order approximation which correctly captures the essential physics.  If
a more precise picture of the system is available, we can improve the
model and get a better approximation to $t_{\text{mfp}}$, in principle.
For strongly coupled plasmas, unfortunately, we do not have such a more
precise picture.  Still, the relations \eqref{jxcu15Sep09},
\eqref{jvmx8Apr09} must give the qualitatively correct time scale
$t_{\text{mfp}}$.

With the above caveats in mind, we will use above relations
\eqref{jxcu15Sep09}, \eqref{jvmx8Apr09} to read off $t_{\text{mfp}}$ for
the Brownian particle in CFT plasma using the bulk Brownian motion.

\subsection{Non-Gaussian random force and Langevin equation}

In the above, we argued that the time scale $t_{\text{mfp}}$ that
characterizes the statistical properties of the random force $R$ is
related to the nontrivial part (connected part) of the 4-point function
of $R$.  Namely, it is related to the non-Gaussianity of the random
force.  Here, let us briefly discuss the relation between
non-Gaussianity and the non-linear Langevin equation.

In subsection \ref{subsec:bndy_BM}, we discussed the linear Langevin
equation \eqref{genLE} for which the friction is proportional to the
momentum $p$.  In other words, the friction coefficient $\gamma(t)$ did
not contain $p$. Furthermore, the random force $R$ was assumed to be
Gaussian.
In many real systems, Gaussian statistics for the random force gives a
good approximation, and the linear Langevin equation provides a useful
approach to study the systems.  However, this idealized physical
situation does not describe nature in general.  For example, even the
simplest case of a Brownian particle interacting with the molecules of
a solvent is rather thought to obey a Poissonian than a Gaussian
statistics (just like the simple model discussed in subsection
\ref{subsec:simple_model}).  It is only in the weak collision limit
where energy transfer is relatively small compared to the energy of the
system that the central limit theorem says that the statistics can be
approximated as Gaussian \cite{Hanggi:1990zz, DHG}.
Furthermore, due to the non-linear fluctuation-dissipation relations
\cite{NLFD}, the non-Gaussianity of random force and the non-linearity
of friction are closely related.
An extension of the phenomenological Langevin equation that incorporates
such non-linear and non-Gaussian situations is an issue that has not yet
been completely settled (for a recent discussion, see \cite{DHG}).

However, the relation between time scales $t_{\text{coll}}$,
$t_{\text{mfp}}$ and $R$ correlators derived in subsection
\ref{subsec:simple_model} does not depend on the existence of such an
extension of the Langevin equation.  Below, we will compute $R$
correlators using the AdS/CFT correspondence and derive expressions for
the time scale $t_{\text{mfp}}$, but that derivation will not depend on
the existence of an extended Langevin equation either.\footnote{More
precisely, the computation in subsection \ref{subsec:holo_ren} is
independent of the existence of any Langevin equation, because we
directly compute the $R$ correlators using the fact that the total force
$F$ equals $R$ in the $m\to\infty$ limit.  On the other hand, in
subsection \ref{subsec:thermalFT_on_ws}, we compute the $R$ correlators
directly, but use the relation \eqref{x_and_R} derived from the linear
Langevin equation.  So, the latter computation is assuming that a
Langevin equation exists at least to the linear order.
}
It would be interesting to use the concrete AdS/CFT setup for Brownian
motion to investigate the above issue of a non-linear non-Gaussian
Langevin equation.  We leave it for future research.

\section{Holographic computation of the $R$-correlator}
\label{sec:holoR4}

In the last section, we saw that $t_{\text{mfp}}$ can be read off if we
know the low-frequency limit of the 2- and 4-point functions of the
random force.  For the connected 4-point function to be nonvanishing, we
need more than the quadratic term $S_0$ in \eqref{S_0} or \eqref{Gen NG
expansion}. Such terms will arise if we keep higher order terms in the
expansion of the Nambu--Goto action.  This amounts to taking into
account the relativistic correction to the motion of the ``cloud''
around the quark mentioned in subsection \ref{subsec:bulk_BM}.  In the
case of the neutral black holes discussed in subsection
\ref{subsec:bulk_BM}, if we keep up to quartic terms (and drop a
constant), the action becomes
\begin{align}
 S&=S_0+S_{\text{int}},\label{S_0+S_int}\\
 S_{\text{int}}&={1\over 16\pi\ap}\int dt\,dr
 \biggl(
 {\dot X^2\over f}-{r^4f}X'^2
 \biggr)^2,
\label{S_int}
\end{align}
where the quadratic (free) part $S_0$ is as given before in \eqref{S_0}.

There are two ways to compute correlation functions in the presence of
the quartic term \eqref{S_int}.  The first one, which is perhaps more
intuitive, is to regard the theory with the action $S_0+S_{\text{int}}$ as a
field theory of the worldsheet field $X$ at temperature $T$ and compute
the $X$ correlators using the standard technique of thermal field theory
\cite{LeBellac}.
The second one, which is perhaps more rigorous but technically more
involved, is to use the GKPW prescription \cite{Gubser:1998bc,
Witten:1998qj} and holographic renormalization \cite{Skenderis:2002wp}
to compute the correlator for the force acting on the boundary Brownian
particle.

The two approaches give essentially the same result in the end, as they
should.  In the following, we will first describe the first approach and
then briefly discuss the the second approach, relegating the technical
details to Appendix \ref{app:holo_ren}\@.  In this section and the next,
for the simplicity of presentation, we will focus on the neutral black
holes of subsection \ref{subsec:bulk_BM}.

\subsection{Thermal field theory on the worldsheet}
\label{subsec:thermalFT_on_ws}

The Brownian string we are considering is immersed in a black hole
background which has temperature $T$ given by \eqref{Hawk_temp_d}.
Therefore, we can think of the string described by the action
\eqref{S_0+S_int} just as a field theory of $X(t,r)$ at temperature $T$,
for which the standard thermal perturbation theory (see {\it e.g.}\
\cite{LeBellac}) is applicable.

For the thermal field theory described by \eqref{S_0+S_int}, let us
compute the real-time connected 4-point function
\begin{align}
 G^{(x)}_{\text{conn}}(t_1,t_2,t_3,t_4)
 &= \ev{\CT[x(t_1)x(t_2)x(t_3)x(t_4)]}_{\text{conn}}\notag\\
 &= \ev{\CT[X(t_1,r_c)X(t_2,r_c)X(t_3,r_c)X(t_4,r_c)]}_{\text{conn}},\label{dlx7Oct09}
\end{align}
where $\CT$ is the time ordering operator and $x(t)=X(t,r_c)$ is the
position of the boundary Brownian particle. In the absence of external
force, $K(\omega)=0$, \eqref{admittance} relates $x$ and random force
$R$ by
\begin{align}
 R(\omega)=-{im\omega x(\omega)\over \mu(\omega)}.\label{x_and_R}
\end{align}
Therefore, using the low-frequency expression for $\mu(\omega)$ given in
\eqref{mu_gen}, we can compute the 4-point function of $R$ from the one for
$x$ in \eqref{dlx7Oct09}.


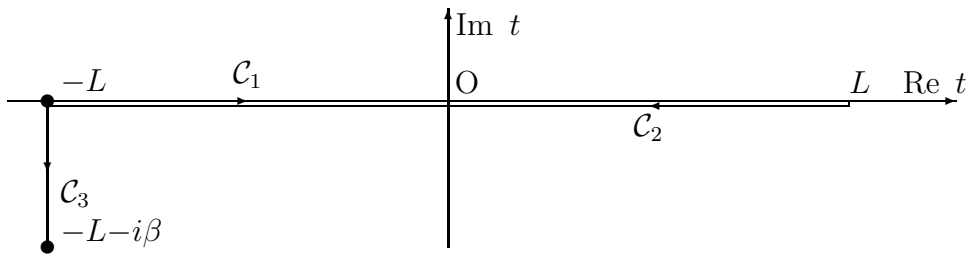
\begin{figure}[ht]
\begin{center}
\begin{picture}(350,100)(-15,-5)
\put(0,55){\circle*{5}}
\put(-15,55){\line(1,0){15}}
\put(0,55){\vector(1,0){75}}
\put(75,55){\line(1,0){75}}
\put(150,55){\line(1,0){75}}
\put(225,55){\line(1,0){75}}
\put(300,55){\vector(1,0){40}}

\put(150,0){\vector(0,1){90}}

\put(300,55){\line(0,-1){2}}

\put(300,53){\vector(-1,0){75}}
\put(225,53){\line(-1,0){75}}

\put(150,53){\line(-1,0){150}}
\put(0,53){\vector(0,-1){25}}

\put(0,28){\line(0,-1){28}}
\put(0,0){\circle*{5}}
\put(5,58){\makebox(0,0)[bl]{$-L$}}
\put(300,58){\makebox(0,0)[bl]{$L$}}
\put(320,58){\makebox(0,0)[bl]{$\Re~t$}}
\put(75,60){\makebox(0,0)[b]{$\CC_1$}}
\put(157,58){\makebox(0,0)[b]{O}}
\put(165,80){\makebox(0,0)[b]{$\Im~t$}}
\put(225,40){\makebox(0,0)[b]{$\CC_2$}}
\put(5,15){\makebox(0,0)[bl]{$\CC_3$}}
\put(5,0){\makebox(0,0)[bl]{$-L{-}i\beta$}}

\end{picture}
\end{center}
\caption{The contour for computing
real-time correlators at finite temperature.}
\label{fig:tcontour}
\end{figure}

As is standard, we can compute such real-time correlators at finite
temperature $T$ by analytically continuing the time $t$ to a complex
time $z$ and performing path integration on the complex $z$ plane along
the contour $C=C_1+C_2+C_3$, where $C_i$ are oriented intervals
\begin{align}
 C_1=[-L,L],\qquad
 C_2=[L,-L],\qquad
 C_3=[-L,-L-i\beta]\label{C123}
\end{align}
as shown in Figure \ref{fig:tcontour}. $L$ is a large positive number
which is sent to infinity at the end of computation.  
We can parametrize the contour $C$ by a real parameter $\lambda$ which
increases along $C$ as
\begin{align}
\begin{array}{rll}
 C_1:&\qquad  z=\lambda-L          &\qquad (0\le\lambda\le 2L) \\
 C_2:&\qquad  z=3L-\lambda         &\qquad (2L\le\lambda\le 4L) \\
 C_3:&\qquad  z=-L+i(4L-\lambda)   &\qquad (4L\le\lambda\le 4L+\beta)
\end{array}
\end{align}
The field $X$ is defined for all values of $\lambda$. Another
convenient parametrization of $C$ is
\begin{align}
\begin{array}{rll}
 C_1:&\qquad z=t,  &\qquad (-L\le t\le L),\\
 C_2:&\qquad z=t, &\qquad  (-L\le t\le L),\\
 C_3:&\qquad z=-L-i\tau,    &\qquad (0\le \tau\le \beta).
\end{array}
\label{paramC1C2C3}
\end{align}
We will denote by $X_{[i]}$ ($i=1,2,3$) the field $X$ on the segment
$C_i$ parametrized by $t$ and $\tau$ in \eqref{paramC1C2C3}.
Henceforth, we will use the subscript $[i]$ for 
a quantity  associated with $C_i$.

The path integral is now performed over $X_{[1]}(t)$, $X_{[2]}(t)$, and
$X_{[3]}(\tau)$, but in the $L\to\infty$ limit the path integral over
$X_{[3]}$ factorizes and can be dropped \cite{LeBellac}.  Therefore,
with the parametrization \eqref{paramC1C2C3}, the path integral becomes
\begin{align}
 \int \CD X\, e^{iS}
 \to \int \CD X_{[1]}\,\CD X_{[2]}\, e^{i(S_{[1]}-S_{[2]})},\label{juii4Oct09}
\end{align}
where $S_{[i]}$, $i=1,2$ are obtained by replacing $X$ with $X_{[i]}$ in
\eqref{S_0+S_int}. The negative sign in front of $S_{[2]}$ in
\eqref{juii4Oct09} is because the direction of the parameter $t$ we took
in \eqref{paramC1C2C3} is opposite to that of $C_2$.

The correlator \eqref{dlx7Oct09} can be written as
\begin{align}
 G^{(x)}_{\text{conn}}(t_1,t_2,t_3,t_4)=
 \ev{\CT_C[X_{[1]}(t_1,r_c)X_{[1]}(t_2,r_c)X_{[1]}(t_3,r_c)X_{[1]}(t_4,r_c)]}_{\text{conn}},\label{kkuk6Jun13}
\end{align}
where $\CT_C$ is ordering along $C$ (in other words, with respect to the
parameter $\lambda$), and can be computed in perturbation theory by
treating $S_0$ as the free part and $S_{\text{int}}$ as an interaction. In doing
that, we have to take into account both the type-1 fields $X_{[1]}$ and
the type-2 fields $X_{[2]}$.  Namely, we have to introduce propagators
not just for $X_{[1]}$ but also between $X_{[1]}$ and $X_{[2]}$ as
follows
\begin{align}
\begin{split}
  D_{[11]}(t-t',r,r')&=\ev{\CT_C[X_{[1]}(t,r)X_{[1]}(t',r')]}_0=\ev{\CT[X(t,r)X(t',r')]}_0
 =D_F(t-t',r,r'),\label{D11D21}\\
 D_{[21]}(t-t',r,r')&=\ev{\CT_C[X_{[2]}(t,r)X_{[1]}(t',r')]}_0 =\ev{X(t,r)X(t',r')]}_0
 =D_W(t-t',r,r').
\end{split}
\end{align}
Here, $\ev{\,}_0$ is the expectation value for the free theory with
action $S_0$ at temperature $T$.  We see that the propagators $D_{[11]}$
and $D_{[21]}$ are equal, respectively, to the usual time-ordered (Feynman)
propagator $D_F$ and the Wightman propagator $D_W$ of the field
$X(t,r)$.  We must also remember that we have not only interaction
vertices that come from $S^{\text{int}}_{[1]}$ and involve $X_{[1]}$,
but also ones that come from $S^{\text{int}}_{[2]}$ and involve $X_{[2]}$.
The second type of vertices come with an extra minus sign.

Using the propagators \eqref{D11D21}, the connected 4-point function is
evaluated, at leading order in perturbation theory, to be
\begin{align}
 &G^{(x)}_{\text{conn}}(\omega_1,\omega_2,\omega_3,\omega_4)
 =
 {i\over 16\pi\ap}
 2\pi \delta(\omega_1+\cdots+\omega_4)
 \int_{r_H}^{r_c} dr\,
 \notag\\
 &\times
 \Biggl\{
 \sum_{\text{perm}\atop (ijkl)}
 \Biggl[ {\omega_i\omega_j\over f} D_{[11]}(\omega_i)D_{[11]}(\omega_j)
 +{r^4f}  \partial_r D_{[11]}(\omega_i)\partial_r D_{[11]}(\omega_j) \Biggr]
 \notag\\
 &\qquad\qquad
 \times\Biggl[ {\omega_k\omega_l\over f} D_{[11]}(\omega_k)D_{[11]}(\omega_l)
 +{r^4f}  \partial_r D_{[11]}(\omega_k)\partial_r D_{[11]}(\omega_l) \Biggr]
 -(D_{[11]}\to D_{[21]})
 \Biggr\}.\label{G_conn_expl}
\end{align}
Here, we wrote down the result in the Fourier space and used a shorthand
notation $D_{[11]}(\omega_i)\equiv D_{[11]}(\omega_i,r,r_c)$.  The
summation is over permutations $(ijkl)$ of $(1234)$.

We are interested in the low frequency limit of this correlator.  In that
limit, the propagators simplify and can be explicitly written down.  In
Appendix \ref{sec:lowE_propagators}, we study the low-frequency
propagators, and the resulting expressions are
\begin{align}
\begin{split}
   D_{[11]}(\omega,r,r_c)&=
   D_F(\omega,r,r_c)=
 {2\pi \ap\over r_H^2}\left[
 {e^{i\omega r_*}+e^{-i\omega r_*}
 \over \omega(1-e^{-\beta\omega})}
 - {e^{i\omega r_*}\over \omega}
 \right],\\[1ex]
 D_{[21]}(\omega,r,r_c)&=
 D_W(\omega,r,r_c)=
 {2\pi \ap\over r_H^2}
 {e^{i\omega r_*}+e^{-i\omega r_*}
 \over \omega(1-e^{-\beta\omega})},
\end{split} 
\label{lowE_D_F,W}
\end{align}
where $r_*$ is the tortoise coordinate introduced in
\eqref{tortoise_gen}.  As explained in \eqref{low_freq_lim_app}, the
precise low frequency limit we are taking is
\begin{align}
 \omega_i\to 0, \qquad\qquad
 \beta,~\omega_i r_*:~\text{fixed}.
 \label{low_freq_lim}
\end{align}
The reason why we have to keep $\omega_i r_*$ fixed is that, no matter
how small $\omega_i$ is, we can consider a region very close to the
horizon ($r_*=-\infty$) such that $\omega_i r_*=\CO(1)$.
If we insert the expressions \eqref{lowE_D_F,W} into \eqref{G_conn_expl}
and keep the leading term in the small $\omega_i$ expansion in the sense
of \eqref{low_freq_lim}, we obtain
\begin{align}
G^{(x)}_{\text{conn}}(\omega_1,\omega_2,\omega_3,\omega_4)
 &\sim
 {i\ap^3 \beta^5\over \omega_1\omega_2\omega_3\omega_4}
  \delta(\omega_1+\cdots+\omega_4)\notag
 \\ 
&\qquad\times
 \sum_{1\le i<j\le 4}(\omega_i+\omega_j)\int_{-\infty}^{0} dr_*\,{r^2\over f}
 e^{-2i(\omega_i+\omega_j)r_*}
 +\CO(\omega^{-2}),\label{X4_expl_naive}
\end{align}
where we ignored numerical factors.  Using \eqref{x_and_R} and
\eqref{mu_gen}, we can finally derive the expression for the $R$
correlator:
\begin{align}
G^{(R)}_{\text{conn}}(\omega_1,\omega_2,\omega_3,\omega_4)
 &\sim
 {i\over \ap\beta^3}
  \delta(\omega_1+\cdots+\omega_4)\notag
 \\ 
&\qquad\times
 \sum_{1\le i<j\le 4}(\omega_i+\omega_j)\int_{-\infty}^{0} dr_*\,{r^2\over f}
 e^{-2i(\omega_i+\omega_j)r_*}
 +\CO(\omega^{2}).\label{R4_expl_naive}
\end{align}

Let us look at the IR part of \eqref{X4_expl_naive}, namely the
contribution from the near-horizon region (large negative
$r_*$). Because $f\sim (d-1)e^{(d-1)r_H r_*}$ near the horizon, the
$r_*$ integral in \eqref{X4_expl_naive} is
\begin{align}
 \int_{-\infty}^{0} dr_*\,{r^2\over f} e^{-2i(\omega_i+\omega_j)r_*}
 \sim
 {r_H^2\over d-1} \int_{-\infty} dr_*\,
 e^{-(d-1)r_H^{}
 r_*}e^{-2i(\omega_i+\omega_j)r_*}
\end{align}
which diverges because of the contribution from the near horizon region,
$r_*\to -\infty$.  We will discuss the nature of this IR divergence
later.

\subsection{Holographic renormalization}
\label{subsec:holo_ren}

Next, let us discuss another way to compute the correlators of the
boundary Brownian motion, following the standard GKPW procedure
\cite{Gubser:1998bc, Witten:1998qj}.  For this approach, we send the UV
cutoff $r_c\to \infty$ and let the string extend all the way to the
AdS boundary $r=\infty$.  The boundary value of $X(t,r)$ is the position
of the boundary Brownian particle: $x(t)=X(t,r\to\infty)$.  The boundary
operator dual to the bulk field $X(t,r)$ is $F(t)$, the total force
(friction plus random force) acting on the boundary Brownian
particle.  The AdS/CFT dictionary
\begin{align}
 \Ev{e^{i\int dt\, F(t)x(t) }}_{\text{CFT}}=e^{iS_{\text{bulk}}[x(t)]}
\end{align}
says that, to compute boundary correlators for $F$, we should consider
bulk configurations for which $X(t,r)$ asymptotes to a given function
$x(t)$ at $r=\infty$, evaluate the bulk action, and functionally
differentiate the result with respect to $x(t)$.  Note that, in the
limit $r_c\to\infty$ or $m\to\infty$ that we take, friction is ignorable
as compared to random force $R$, and $F$ correlators are the same as $R$
correlators \cite{CasalderreySolana:2006rq}. Roughly speaking, because
the Brownian particle does not move in the $m\to\infty$ limit, there
will be no friction and thus $R=F$.

In the end, the resulting 4-point function $\ev{FFFF}$ is essentially
given by the interaction term in the action, with the $X$ fields replaced
by the boundary-bulk propagators. Namely,
\begin{align}
 \ev{\CT[F(t_1)F(t_2)F(t_3)F(t_4)]}
 &\sim
 {1\over 16\pi \ap}\int dt\,dr\sum_{\text{perm} \atop (ijkl)}
 \left[ -{\partial_t K(t_i)\, \partial_t K(t_j)\over f}
   +{r^4f}\,\partial_r K(t_i)\, \partial_r K(t_j)\right]\notag\\
 &\qquad\times
 \left[ -{\partial_t K(t_k)\, \partial_t K(t_l)\over f}
   +{r^4f}\,\partial_r K(t_k)\, \partial_r K(t_l) \right],\label{GKPW_naive}
\end{align}
where $K(t_i)\equiv K(t,r|t_i)$ is the boundary-bulk propagator from the
boundary point $t_i$ to the bulk point $(t,r)$.  This is the Witten
diagram rule that we naively expect. However, because the worldvolume
theory of a string is different from, {\it e.g.}\ a Klein--Gordon
scalar, a careful consideration of holographic renormalization (see
\cite{Skenderis:2002wp} for a review) is necessary. Indeed, the naive
expression is \eqref{GKPW_naive} is UV divergent and needs
regularization. Furthermore, our black hole spacetime is a Lorentzian
geometry and we should apply the rules of Lorentzian AdS/CFT
\cite{Skenderis:2008dg, vanRees:2009rw}.  As is explained in Appendix
\ref{app:holo_ren}, after all the dust has been settled, the $F$
correlator gives exactly the same IR divergence as the naive computation
of the $R$ correlator, \eqref{R4_expl_naive}.  This implies that this IR
divergence we are finding is not an artifact but a real thing to be
interpreted physically.\footnote{Although the IR parts are the same, the
result obtained in the previous subsection \ref{subsec:thermalFT_on_ws}
using the worldsheet thermal field theory is not quite the same as the
one obtained in this subsection \ref{subsec:holo_ren} using holographic
renormalization, due to the counter terms added to the latter at the UV
cutoff $r=r_c$.}

It is worth pointing out that the result \eqref{GKPW_naive} has a
similar structure to the one we saw in the toy model \eqref{jtet8Apr09},
with the propagator $K(t)$ roughly corresponding to $P(t)$.  It would be
interesting to find an improved toy model which precisely reproduces the
structure \eqref{GKPW_naive}.

In Appendix \ref{apss:retarded4pt}, we also computed the \emph{retarded}
4-point function of random force.  The expression is free from both IR
and UV divergences and the final result is finite.  However, because we
do not know how to relate the retarded 4-point function and
$t_{\text{mfp}}$, this cannot be used to compute $t_{\text{mfp}}$.  It
would be interesting to find a microscopic model that directly relates
retarded correlators and $t_{\text{mfp}}$.

\subsection{General polarizations}

The argument so far has been as if there were only one field $X$ and the
associated random force $R$. However, in the general $d>3$ case we have
$n=d-2>1 $ fields $X^I$, $I=1,2,\dots,n$.  Considering all $X^I$, the bulk
action \eqref{S_int} actually becomes
\begin{align}
 S_{\text{int}}&={1\over 16\pi\ap}\int dt\,dr
 \biggl[
 {(\dot X^I)^2\over f}-{r^4f}\,(X^I{}')^2
 \biggr]^2.
\end{align}
The associated random force $R^I$ has $n$ components too.

The computation of 4-point functions in this multi-component case can be
done completely in parallel with the one-component case.  Let us define
\begin{align}
 G^{(x)IJKL}_{\text{conn}}(t_1,t_2,t_3,t_4)\equiv
\ev{\CT[X^I(t_1,r_c)X^J(t_2,r_c)X^K(t_3,r_c)X^L(t_4,r_c)]}.
\end{align}
This is nonvanishing only if some indices are identical.  More
precisely, the only nonvanishing cases are (i) all indices are
identical, $I=J=K=L$, or (ii) indices are pairwise identical, $I=J\neq
K=L$, $I=K\neq J=L$, or $I=L\neq J=K$.  

In case (i), the resulting 4-point function is exactly the same as the
one-component case \eqref{G_conn_expl}.  Consequently, the IR form of
the random force correlator $G^{(R)IIII}_{\text{conn}}$ is the same as
the one-component case \eqref{R4_expl_naive}.

In case (ii), on the other hand, the 4-point function becomes
\begin{align}
 &G^{(x)IIJJ}_{\text{conn}}(\omega_1,\omega_2,\omega_3,\omega_4)
 =
 {i\over 16\pi\ap}
 2\pi \delta(\omega_1+\cdots+\omega_4)
 \int_{r_H}^{r_c} dr\,
 \notag\\
 &\times
 \Biggl\{
 8
 \Biggl[ {\omega_1\omega_2\over f} D_{[11]}(\omega_1)D_{[11]}(\omega_2)
 +{r^4f}  \partial_r D_{[11]}(\omega_1)\partial_r D_{[11]}(\omega_2) \Biggr]
 \notag\\
 &\qquad\qquad
 \times\Biggl[ {\omega_3\omega_4\over f} D_{[11]}(\omega_3)D_{[11]}(\omega_4)
 +{r^4f}  \partial_r D_{[11]}(\omega_3)\partial_r D_{[11]}(\omega_4) \Biggr]
 -(D_{[11]}\to D_{[21]})
 \Biggr\}.\label{G_conn_expl_IIJJ}
\end{align}
The IR form of the random force correlator is
\begin{align}
G^{(R)IIJJ}_{\text{conn}}(\omega_1,\omega_2,\omega_3,\omega_4)
 &\sim
 {i\over \ap\beta^3}
  \delta(\omega_1+\cdots+\omega_4)\,\int_{-\infty}^{0} dr_*\,{r^2\over f}
 \notag
 \\ 
&\qquad\times
 \sum_{1\le i\le 2,\, 3\le j\le 4}(\omega_i+\omega_j)
 e^{-2i(\omega_i+\omega_j)r_*}
 +\CO(\omega^{2}).\label{R4_expl_naive_IIJJ}
\end{align}
Comparing this with the expectation from the field theory side,
\eqref{RIRJRKRLconn} we observe the same structure. Namely, the
connected 4-point functions are nonvanishing only when the polarization
indices are all or pairwise identical.  The precise relative values of
the nonvanishing 4-point functions are model-dependent and not important;
in the simple model of subsection \ref{subsec:simple_model}, it depends
on our choice of the expectation values \eqref{ev_ee,eeee}.  

\subsection{Comment on the basis}
\label{ss:comment_basis}

In this section, we computed the correlation functions for type-1 fields
$X_{[1]}$, such as \eqref{kkuk6Jun13}, as the quantities to be matched
with those in the simple model presented in subsection
\ref{subsec:simple_model}.  One may wonder whether it is more
appropriate to use correlation functions in some other basis, such as
the retarded/advanced ($r$-$a$) basis \cite{Chou:1984es,
Aurenche:1991hi, Aurenche:1993vt}.  For example, $G_{rrrr}$ in the
$r$-$a$ basis has no knowledge of time ordering unlike $G_{[1111]}$ in
the 1-2 basis and might seem more natural quantity to consider.
However, recall that the analysis in subsection
\ref{subsec:simple_model} is a classical one; therefore, the difference
between $G_{[1111]}$ and $G_{rrrr}$ is quantum and thus negligible in
our approximation.  Clearly, $G_{[1111]}$ is much easier to compute than
$G_{rrrr}$ and we will use the former to extract $t_{\rm mfp}$
below.\footnote{We have checked that $G_{[1111]}$ and $G_{rrrr}$ indeed
give the same result.}

\section{The IR divergence}
\label{sec:IRdiv}

In the last section, we computed the connected 4-point function for the
random force $R$ and found that the low-frequency expression,
\begin{align}
G^{(R)}_{\text{conn}}(\omega_1,\omega_2,\omega_3,\omega_4)
 &\sim
 {i\over \ap\beta^3}
  \delta(\omega_1+\cdots+\omega_4)\notag
 \\ 
&\qquad\times
 \sum_{1\le i<j\le 4}(\omega_i+\omega_j)\int_{-\infty}^{0} dr_*\,{r^2\over f}
 e^{-2i(\omega_i+\omega_j)r_*},
 \label{R4_IRdiv}
\end{align}
has an IR divergence coming from the integral in the near horizon
region.  What is the physical reason for this divergence?  Very near the
horizon, the expansion of the Nambu--Goto action in the transverse
fluctuation $X$ breaks down because the proper temperature becomes
higher and higher as one approaches the horizon and, as a result, the string
fluctuation gets wilder and wilder.  The correct thing to do in
principle is to consider the full non-linear Nambu--Goto action, but this
is technically very difficult.
Instead, a physically reasonable estimate of the result is the
following.  Let us introduce an IR cutoff near the horizon at
\begin{align}
 r_{\rm s}=r_H+\epsilon,
\label{IRcutoff}
\end{align}
where $\epsilon \ll r_H$.  We take this cutoff $r_{\rm s}$ to be the
radius where the expansion of the Nambu--Goto action becomes bad.  Then,
in IR-divergent expressions such as \eqref{R4_expl_naive}, we simply
throw away the contribution from the region $r< r<r_{\rm s}$ by taking
the integral to be only over $r>r_{\rm s}$.  Of course, to obtain a more precise
result, we should include the contribution from the region
$r_H<r<r_{\rm s}$ with the higher order terms in the expansion of the
Nambu--Goto action taken into account.  However, we expect that the
contribution from this region $r_H<r<r_{\rm s}$ will be of the same
order as the contribution from the region $r>r_{\rm s}$ and, therefore,
we can estimate the full result by just keeping the latter contribution.

With this physical expectation in mind, let us evaluate the
mean-free-path time $t_{\text{mfp}}$ by introducing the IR cutoff
\eqref{IRcutoff}.  The parameter $\epsilon$ appearing in
\eqref{IRcutoff} can be related to the proper distance from the horizon,
$s$, as follows:
\begin{align}
 s= \int_{r_H}^{r_H+\epsilon} {dr\over r\sqrt{f}}
 \sim \int_{r_H}^{r_H+\epsilon} {dr\over \sqrt{(d-1)r_H(r-r_H)}}
 =\sqrt{2\epsilon\over (d-1)r_H}.
\end{align}
Therefore
\begin{align}
 \epsilon
 \sim s^2 r_H,
\end{align}
where we dropped numerical factors.  In the tortoise coordinate $r_{*}$,
the cutoff is at
\begin{align}
 r_{*}^{\rm s} \sim -{1\over (d-1)r_H} \log s^2,
 \label{r*_ito_s}
\end{align}
where we used \eqref{tortoise_NH}.

The introduction of an IR cutoff of the geometry near the horizon also
means that the resulting expressions such as \eqref{R4_IRdiv}, with the
IR cutoff imposed, is valid only for frequencies larger than a certain
cutoff frequency $\omega_{\rm s}$.  We can relate $\omega_{\rm s}$ with
the geometric cutoff $r_*^{\rm s}$ as follows.  If we cut off the
geometry at $r_*=r_*^{\rm s}$, we have to impose some boundary condition
there (just as for the brick wall model).  For example, let us impose a
Neumann boundary condition.  As was shown in \eqref{lowfreq_f,g}, for
very low frequencies, the solutions to the wave equation behave as
\begin{align}
 f_\omega(r)\sim e^{i\omega r_*}+e^{-i\omega r_*}.
\end{align}
For this to satisfy Neumann boundary condition
$\partial_{r_*}f_\omega(r)|_{r_*=r_*^{\rm s}}=0$, we need
$\omega={n\pi/ r_*^{\rm s}}$ where $n\in\bbZ$.  Namely, the frequency
has been discretized in units of $\pi/ |r_*^{\rm s}|$.  Therefore, the
smallest possible frequency is
\begin{align}
 \omega_{\rm s}
 \sim {1\over |r_*^s|}
 \sim {1\over \beta\log(1/s)}.
 \label{omegas_ito_s}
\end{align}

If we use \eqref{r*_ito_s} and \eqref{omegas_ito_s}, the correlator
\eqref{R4_IRdiv} becomes
\begin{align}
G^{(R)}_{\text{conn}}(\omega_1,\omega_2,\omega_3,\omega_4)
 &\sim {i\over\ap\beta^3} \delta(\omega_1+\cdots+\omega_4)\,
 \omega_{\rm s}r_H^2\int_{r_*^{\rm s}} dr_* e^{-(d-1)r_H r_*}
 \notag\\
 &\sim {i\over\ap\beta^3} \delta(\omega_1+\cdots+\omega_4)\,
 \omega_{\rm s}r_H^2\,
 {e^{-(d-1)r_H r_*^{\rm s}} \over r_H }\notag\\
 &\sim
 {is^2 \omega_{\rm s}\over \ap \beta^4}\delta(\omega_1+\cdots+\omega_4)
 \sim
 {is^2 \over \ap \beta^5 \log(1/s)}\delta(\omega_1+\cdots+\omega_4).
\end{align}
On the other hand, from \eqref{RRcorr}, the 2-point function is
\begin{align}
 G^{(R)}(\omega_1,\omega_2)
 &\sim {1\over \ap\beta^3}\delta(\omega_1+\omega_2)
\end{align}
Comparing above results and the toy model results \eqref{jxcu15Sep09},
\eqref{jvmx8Apr09}, we obtain
\begin{align}
 t_{\rm mfp}\sim {1\over \mu}\sim {\ap\beta\over s^2\log(1/s)},\qquad
 P(\omega=0)\sim {1\over \beta s\sqrt{\log(1/s)}}.
 \label{tmfp_ft0_0}
\end{align}

Now the question is how to determine the length $s$.  This must be the
place where the expansion \eqref{S_0+S_int} of the Nambu--Goto action
becomes bad.  One can show that this occurs a proper length $\sim\sqrt{\ap}$
away from the horizon due to thermal fluctuation (Hawking radiation) in
the black hole background (for an argument in more general setups see
subsection \ref{subsec:tmfp_gen}).  This leads us to set
\begin{align}
 s\sim \sqrt{\ap}.\label{s_ansatz}
\end{align}
At this point, the local proper temperature becomes of the order of the
Hagedorn temperature, $\sim 1/\sqrt{\ap}$.  The above condition must be
the same as the condition that the loop correction of the worldsheet
theory to the 4-point function $\ev{F^4}$ becomes of the same order as
the tree level contribution.

If we substitute \eqref{s_ansatz} into \eqref{tmfp_ft0_0}, we obtain
\begin{align}
 t_{\rm mfp}
 \sim {1 \over T\log \lambda},\qquad
 P(\omega=0)\sim {T\lambda^{1/4}\over \sqrt{\,\log\lambda}}
 \label{tmfp_ft0}
\end{align}
where, following the convention of the $d=5$ (AdS$_5$) case, we defined
the ``'t Hooft coupling'' by
\begin{align}
 \lambda\equiv {l^4\over \ap^2},\label{'tHooft_def}
\end{align}
where we restored the AdS radius $l$ which we have been setting to one.

The result \eqref{tmfp_ft0} is quite interesting.  In
\cite{deBoer:2008gu}, the collision duration time $t_{\text{coll}}$ was
determined to be
\begin{align}
 t_{\text{coll}}\sim{1\over T}.\label{tcoll_ft0}
\end{align}
Therefore, $t_{\text{mfp}}$ given in \eqref{tmfp_ft0} implies that a
plasma particle can be thought of as in interaction with roughly
$\log\lambda$ other particles simultaneously.

Even if we take into account the fact that $X^I$ has in general more
than one component $(I=1,2,\dots,n=d-2)$ and use the results such as
\eqref{RIRJRKRLconn}, \eqref{R4_expl_naive_IIJJ}, we end up the same
estimate for $t_{\text{mfp}}$ as far as its order is concerned.

\section{Generalizations}
\label{sec:gen's}

In the previous section, we derived using AdS/CFT the expression for the
mean-free-path time $t_{\text{mfp}}$ for the simple case of neutral
plasma.  In this section, we sketch how this generalizes to the more
general metric \eqref{gen non-rotating metric} and present the
expression for the mean-free-path time for more general systems such as
charged plasmas.  As an example, we will apply the result to the STU
black hole.

\subsection{Mean-free-path time for the general case}
\label{subsec:tmfp_gen}

We are interested in computing the mean-free-path time in field theory
by analyzing the motion of a Brownian string in the metric \eqref{gen
non-rotating metric}.  For that, as has been explained in section
\ref{sec:time_scales} for the neutral case, we need to compute the
4-point function of the random force in addition to the 2-point function.

Expanding the Nambu--Goto action in the background metric (\ref{gen
non-rotating metric}) up to quartic order, the action for the string in
the tortoise coordinate defined in \eqref{tortoise_def_gen} is given as
follows:
\begin{align}
S&=S_0+S_{\text{int}},\label{action_gen_0+int}
 \\
S_0&= \frac{1}{4\pi\alpha'}\int dt\,dr_*\, G \,
 (\dot{X}^2-X'^2),
 \label{Gen NG 2nd}
 \\
S_{\text{int}}&= \frac{1}{16\pi\alpha'}\int dt\,dr_*\, {G^2\over h_t f}
 (\dot{X}^2-X'^2)^2,
 \label{Gen NG 4th} 
\end{align}
where we dropped a constant independent of the field $X$, and
$\,\dot{}=\partial_t$, $'=\partial_{r_*}$.  As we discussed in
subsection \ref{subsec:holo_ren} for the simple neutral case, we can use
$S_{\text{int}}$ as the interaction term and apply the usual GKPW rule
to compute correlators for the random force\,\footnote{Recall that in this
setup the force $F$ is equal to the random force $R$.} $F$ dual to the
bulk field $X$.
As before, the naive result from the GKPW prescription includes both UV
and IR divergences.  Using holographic renormalization, which is
discussed in Appendix \ref{app:holo_ren} for the neutral case, we can
remove the UV divergence by adding counter terms to the action.  The IR
divergence, on the other hand, signals the breakdown of the quartic
approximation \eqref{action_gen_0+int}.  We regulate this divergence by
introducing an IR cutoff at $r_*=r_*^s$ near to the horizon, whose
physical motivation was explained in section \ref{sec:IRdiv}.

Following the same analysis as in section \ref{sec:IRdiv} now with the
interaction term (\ref{Gen NG 4th}), we obtain an expression similar to
(\ref{R4_expl_naive}) for the connected random force 4-point function.
The dominant contribution comes from the near-horizon region and is
given in frequency space by
\begin{align}
 \ev{\CT[F^4]}_{\text{conn}}
 &\sim
 {i\over\alpha'\beta^3}
 \delta(\omega_1+\cdots+\omega_4)\int_{r^s_*} dr_* {G^2\over f h_t}
\sum\limits_{1\leq i<j\leq 4}(\omega_i+\omega_j)e^{-2i(\omega_i+\omega_j)r_*},\label{F4gen_before_int}
\end{align}
where $r_*^{\rm s}$ is the aforementioned IR cutoff (in the tortoise
coordinate).
Let the IR cutoff in the $r$ coordinate be at $r=r_H+\epsilon\equiv
r_{\rm s}$.  The parameter $\epsilon$ is related to the proper distance $s$
from the horizon as
\begin{align}
 s=\int^{r_H +\epsilon}_{r_H} \sqrt{h_r\over f\,}\,dr
 \approx \sqrt{{2\epsilon\, h_r(r_H)\over k_H}},\qquad
 \epsilon\approx {s^2 k_H\over 2 h_r(r_H)}.\label{Gen proper distance}
\end{align}
Using the relation \eqref{tortoise_def_gen} between $r_{\rm s}$ and
$r_*^{\rm s}$, we can estimate the cut-off integral
\eqref{F4gen_before_int} as
\begin{align}
 \ev{ \mathcal{T}[F^4] }
 &\sim
 {G^{2}(r_H)\,\omega_{\rm s} \over \alpha' s^2}
 \delta(\omega_1+\cdots+\omega_4),
\end{align}
where $\omega_{\rm s}$ is the smallest frequency for which the expansion
\eqref{action_gen_0+int} is valid.  Combining this with the result
\eqref{gen RR correlator} for the 2-point function, 
the mean-free-path time is estimated as
\begin{align}
 t_{\rm mfp}\sim {\alpha'\beta^2 \omega_{\rm s} \over s^2}.
 \label{tmfp_gen_0}
\end{align}

Now, let us determine the IR cutoff parameters $s$ (or equivalently
$\epsilon$) and $\omega_{\rm s}$ appearing in \eqref{tmfp_gen_0}.  As
before, we take the IR cutoff to be the location where $S_0$ and
$S_{\text{int}}$ become of the same order.  As is clear from \eqref{Gen
NG 2nd}, \eqref{Gen NG 4th}, the expansion of the Nambu--Goto action
becomes bad at the location where
\begin{align}
 {G\over h_t f}\dot{X}^2,~
 {G\over h_t f}X'^2\sim 1.
 \label{Gen flux cond}
\end{align}
So, we would like to estimate $\dot{X}$, $X'$.  Near the horizon,
$r\approx r_H$, we can write the action (\ref{Gen NG 2nd}) as
\begin{align}
S_0\sim 
 \frac{1}{2}\int dt\,dr_*  (\dot{\Xt}^2-\Xt'^2),\qquad
\Xt\equiv \sqrt{G(r_H)\over 2\pi\alpha'}\, X.
\end{align}
There being no dimensional quantity in the problem other than the
temperature $T$, we must have $\dot{\Xt}, \Xt'\sim T$, namely $|\dot X|,|X'|
\sim \sqrt{2\pi\alpha'/ G(r_H)\,}\,T$. So, the condition \eqref{Gen flux
cond} determines the IR cutoff to be at
\begin{align}
r-r_H=\epsilon \sim {\alpha' T^2\over k_H h_t(r_H)}.\label{Gen flux cond 2}
\end{align}
In term of $s$, the IR cutoff is at the string length:
\begin{align}
 s\sim \sqrt{\alpha'}.
\end{align}

It is more subtle to determine the parameter $\omega_{\rm s}$.  In
Appendix \ref{sec:lowE_wave} (around Eq.\ \eqref{alpha_omega}), the
following was shown.  Let us we choose the tortoise coordinate $r_*$ to
be related to $r$ near the horizon as
\begin{align}
 r_*\approx {1\over 4\pi T}
 \log\left({r-r_H\over L_H}\right),\label{rs_vs_r_NH_precise}
\end{align}
where $L_H$ is defined through the following integral
\begin{align}
 \int_\infty^r {dr\over fG}\sqrt{h_r\over h_t}
 ={1\over 4\pi G_H T}\log\left({r-r_H\over L_H}\right)+\CO(r-r_H)\label{L_H_def_text}
\end{align}
for $r\approx r_H$.  Then the solution $f_\omega(r)$ to the wave
equation \eqref{gen eqm}, satisfying a normalizable boundary condition
at infinity, will have the form
\begin{align}
 f_\omega(r)\sim e^{i\omega r_*} - e^{-i\omega r_*}
\end{align}
for small $\omega$.  More precisely, we have
\begin{align}
 f_\omega(r)\sim e^{i\omega r_*} - e^{i\alpha_\omega}e^{-i\omega r_*},
\qquad\alpha_\omega=\CO(\omega^2).
 \label{f=e-e}
\end{align}
Now, let us we impose some boundary condition at $r_*^{\rm s}$, such as
a Neumann boundary condition $\partial_{r_*} f_\omega=0$, then the
frequency $\omega$ gets discretized in units of $\Delta
\omega=\pi/|r_*^{\rm s}|$.  Note that, if $\alpha_\omega=\CO(\omega)$ as
$\omega\to 0$, then the coefficient of the $\CO(\omega)$ term will
affect the value of $\Delta \omega$; this is why \eqref{f=e-e} was
important.  This motivates the following choice for the minimum
frequency:
\begin{align}
 \omega_{\rm s}\sim \Delta\omega\sim {1\over |r_*^{\rm s}|}
 \sim {1\over \beta\log\left({L_H\over \epsilon}\right)}
 \sim {1\over \beta\log\left({\beta L_H\over s^2}\sqrt{h_t(r_H) h_r(r_H)}\right)}.
\end{align}

Substituting in the above expressions for $s,\omega_{\rm s}$, the
mean-free-path time \eqref{tmfp_gen_0} is
\begin{align}
 t_{\text{mfp}}
 \sim
 {1\over T\log\bigl(\eta\sqrt{\lambda}\bigr)},
\qquad\qquad
\eta\equiv {L_H\over T}\sqrt{h_t(r_H) h_r(r_H)\,},
\label{tmfp_final} 
\end{align}
where $\lambda$ is the ``'t Hooft coupling'' defined in
\eqref{'tHooft_def}.  Note that the nontrivial effect of charge only
enters through the logarithm and hence the dependence of
$t_{\text{mfp}}$ on it is very mild in the strongly coupled case
$\lambda\gg 1$.

\subsection{Application: STU black hole}
\label{subsec:app_STU}

The AdS/CFT correspondence has been successfully used to extract the
properties of field theory plasmas.  A particularly interesting case is
a 4-dimensional charged plasma, because it is relevant for the
experimentally generated quark-gluon plasma with net baryon charge.
One notable situation to realize 4-dimensional charged plasmas in the
AdS/CFT setup is the spinning D3-brane, which in the decoupling limit
gives $d=4$, $\CN=4$ SYM with nonvanishing $R$-charges.  We can have
three different $R$-charges corresponding three Cartan generators of the
$SU(4)\cong SO(6)$ $R$-symmetry group.
As already mentioned in subsection \ref{subsec:BM_bulk_gen's}, on the
gravity side this corresponds to a Kerr black hole in AdS$_5\times S^5$
with three angular momenta in the $S^5$ directions \cite{Kraus:1998hv,
Russo:1998by}. Upon compactifying on $S^5$, this reduces to the
so-called STU black hole of the five-dimensional supergravity
\cite{Behrndt:1998jd, Cvetic:1999xp}.  From this five-dimensional
perspective, the STU black hole is a non-rotating black hole with three
$U(1)$ charges.
There has been much study \cite{Herzog:2006se, Caceres:2006dj,
Lin:2006au, Avramis:2006ip, Armesto:2006zv, Caceres:2006ta,
Avramis:2006em, Herzog:2007kh, Maeda:2008hn} on the properties of the
$R$-charged field theory plasma using the STU black hole.
Here, we would like to apply the machineries we have developed in the
previous sections to the computation of the mean-free-path time for the
Brownian particle in $R$-charged plasma dual to the STU black hole.

\subsubsection{The STU black hole}

The 10-dimensional metric of the STU black hole is given
by~\cite{Behrndt:1998jd}:\footnote{The horizon of the STU black hole can
be either $S^3$, $\bbR^3$, or $H^3$, but we are focusing on the planar
$\bbR^3$ case, corresponding to a charged plasma in flat $\bbR^3$.}
\begin{align}
 ds_{10}^2&=\sqrt{\Delta}\, ds_5^2 +{l^2\over \sqrt{\Delta}}
 \sum_{i=1}^3 X_i^{-1}
 \left[d\mu_i^2+\mu_i^2\left(d\psi_i+{A^i\over l}\right)^2\right],
 \label{STU gen}
 \\
 ds_5^2&=-\frac{f}{\mathcal{H}^{2/3}}dt^2+\mathcal{H}^{1/3}\left(\frac{dr^2}{f}+r^2(dX^I)^2\right),\nonumber\\
 f(r)&=\frac{r^2}{l^2}\mathcal{H}-\frac{m}{r^2},\qquad
 \CH=H_1H_2H_3,\qquad H_i=1+\frac{q_i}{r^2},\notag\\
 X_i&=H_i^{-1}\CH^{1/3},\qquad A^i=\sqrt{m\over q_i}\,(1-H_i^{-1})dt,\qquad
 \Delta=\sum_{i=1}^3X_i \mu_i^2,\notag\\
 \mu_1&=\sin\theta_1,\qquad
 \mu_2=\cos\theta_1\sin\theta_2,\qquad
 \mu_3=\cos\theta_1\cos\theta_2
 \notag
\end{align}
with $i=1,2,3$. Here, $X^I$, $I=1,2,3$ are spatial directions along the
boundary and $l$ is the AdS radius.  The four parameters $m,q_i$ are
related to the mass and three electric charges of the STU black hole.
It is convenient to introduce the dimensionless quantities
\begin{align}
 \kappa_i={q_i\over r_H^2},\qquad i=1,2,3.
\end{align}
The horizon is at
$r=r_H$ where $r_H$ is the largest solution to $f(r)=0$.  The latter
equation relates $m$ to $r_H$ and $\kappa_i$ as
\begin{align}
 m&={r_H^4\over l^2}\CH(r_H)
 ={r_H^4\over l^2}(1+\kappa_1)(1+\kappa_2)(1+\kappa_3).\label{m_rel}
\end{align}
The Hawking temperature is given by
\begin{align}
 T
 ={r_H\over 2\pi}\,
 {2+\kappa_1+\kappa_2+\kappa_3-\kappa_1\kappa_2\kappa_3
 \over \sqrt{(1+\kappa_1)(1+\kappa_2)(1+\kappa_3)}}.\label{T_STU}
\end{align}
From the five-dimensional point of view, the STU black hole is
electrically charged under the gauge fields $A^i$ and the associated
chemical potentials are
\begin{align}
 \Phi^i={1\over \kappa^2_5}\left[A_t^i(r=\infty)-A_t^i(r=r_H)\right]
 =-{r_H^2\over \kappa_5^2 l}\,
 {\sqrt{\kappa_i}\,\prod_{j=1}^3 (1+\kappa_j)\over 1+\kappa_i}.\label{Phi_STU}
\end{align}
Here $\kappa_5^2=8\pi G_5$ is the five-dimensional Newton constant and
\begin{align}
 G_5
 ={G_{10}\over V_{S^5}}
 ={8\pi^6 g_s^2 \ap^4\over \pi^3 l^5}={\pi l^3\over 2N^2},
\end{align}
where $N$ is the rank of the boundary gauge theory.  For expressions for
other physical quantities, such as energy density, entropy density, and
charge density, see {\it e.g.}\ \cite{Mas:2006dy}.  From thermodynamical
stability, the parameters $\kappa_i$ are restricted to the range
\cite{Cvetic:1999rb}
\begin{align}
 2-\kappa_1-\kappa_2-\kappa_3+\kappa_1\kappa_2\kappa_3>0.
 \label{thermo_stab}
\end{align}

We can shift the gauge potential $A^i$ so that its value on the horizon
is zero:
\begin{align}
 \CA^i(r)\equiv A^i(r)-A^i(r_H).
\end{align}
If we accordingly shift the angular variable by
\begin{align}
 \psit_i\equiv \psi_i+A^i_t(r_H)\,
\end{align}
then the metric \eqref{STU gen} becomes
\begin{align}
 ds_{10}^2=\sqrt{\Delta}\, ds_5^2 +{R^2\over \sqrt{\Delta}}
 \sum_{i=1}^3 X_i^{-1}[d\mu_i^2+\mu_i^2(d\psit_i+\CA^i/R)^2].
\label{STU gen2}
\end{align}

\subsubsection{Background configuration}

We first want to find a background configuration of a string in the 10
dimensional geometry \eqref{STU gen} or \eqref{STU gen2}, so that we can
start expanding the Nambu--Goto action around it.  If we restrict
ourselves to configurations with trivial $\theta_a$ dependence, the
relevant line element can be written as
\begin{align}
 ds^2&=-\alpha\, dt^2+\beta\, dr^2+\gamma (dX^I)^2
 +\sum_{i=1}^3\epsilon_i(d\psit_i+\phi_i dt)^2.\label{STUmetricabc}
\end{align}
Here $\alpha,\beta,\gamma,\epsilon_i,\phi_i$ are functions of $r$ and
$\theta_a$ which can be read off from \eqref{STU gen2}. For example,
$\alpha=\Delta^{1/2}f\CH^{-2/3}$. Parametrize the worldsheet by $t,r$
and take the following ansatz:
\begin{align}
 X^I(t,r)=0,\qquad
 \psit_i(t,r)=\omegat_i t+\varphi_i(r).\label{drag_ansatz}
\end{align}
The string is straight in the AdS$_5$ part of the spacetime.  On the
other hand, the angular momenta in the $S^5$ directions are expected to
drag the string in these directions and $\omegat_i,\varphi_i$ correspond
to nontrivial drifting/trailing of the string \cite{Herzog:2006se,
Caceres:2006dj, Herzog:2007kh}.  The Euler--Lagrange equation for
$\varphi(r)$ states that $\pi^r_{\varphi_i} \equiv {\partial
L_{\text{NG}}/\partial (\partial_r \psit_i)} = {\partial
L_{\text{NG}}/\partial \varphi_i}$ is constant along the string.  The
quantity $\pi^r_{\varphi_i}$ corresponds to the inflow of angular
momenta (or, from the five-dimensional point of view, electric charges)
from the ``flavor D-brane'' at the UV cutoff $r=r_c$, and how to choose
them depends on the physical situation one would like to consider
\cite{Herzog:2007kh}.  Here, let us focus on the case where the string
endpoint on the ``flavor D-brane'' is free and there is no inflow, {\it
i.e.}, $\pi^r_{\varphi_i}=0$.  This corresponds to a boundary Brownian
particle neutral under the $R$-symmetry.  This is physically appropriate
because we want to compute the random force correlators unbiased by the
effects of the charge of the probe itself.  It is not difficult to see
that setting $\pi^r_{\varphi_i}=0$ leads to $\varphi_i=0$ by examining
the Euler--Lagrange equations.

Let us next turn to the angular velocity $\omegat_i$.  Given $\varphi_i=0$,
the induced metric on the worldsheet is
\begin{align}
 ds_{ind}^2=-\alpha dt^2+\beta dr^2
 +\sum_{i=1}^3 \epsilon_i (\omegat_i+\phi_i)^2 dt^2.
\end{align}
The determinant of this induced metric is
\begin{align}
 \det g\propto -\alpha+\sum_{i=1}^3 \epsilon_i(\omegat_i+\phi_i)^2.
\end{align}
This must be always non-positive for the configuration to physically
make sense.  This condition is most stringent at the horizon $r=r_H$
where $\alpha\propto f=0$, $\phi_i=\CA^i_t(r_H)/l=0$. So, we need
\begin{align}
 \sum_i \epsilon_i \omegat_i^2\le 0.
\end{align}
Since $\epsilon_i\ge0$, this means that
\begin{align}
 \omegat_i=0.
\end{align}
Namely, the background configuration is simply
\begin{align}
 X^I(t,r)=\psit_i(t,r)=0.\label{STU_bg_soln}
\end{align}
Note that the angular motion is trivial only in the $\psit_i$
coordinates and in the original $\psi_i$ coordinates there is
non-vanishing angular drift.

So far we have been treating $\theta_a$ as constant. However, this is
not correct and an arbitrary choice of $\theta_a$ will not satisfy the
full equations of motion.  Below, we will consider the following three
cases:
\begin{itemize}
 \item[(i)] 1-charge case: $\kappa_1=\kappa\neq 0$, $\kappa_2=\kappa_3=0;$\quad $\theta_1=\pi/2$,
 \item[(ii)] 2-charge case: $\kappa_1=0$, $\kappa_2=\kappa_3=\kappa\neq 0;$\quad  $\theta_1=0$,
 \item[(iii)] 3-charge case: $\kappa_1=\kappa_2=\kappa_3=\kappa\neq 0;$\quad $\theta_1,\theta_2$: arbitrary.
\end{itemize}
It can be shown \cite{Herzog:2007kh} that the above values of $\theta_a$
are necessary for all the equations of motion to be satisfied. These
values make sense physically since, if the angular momentum around an
axis is nonvanishing, the string wants to orbit along the circle of the
largest possible radius around that axis. This is achieved by the above
choices of $\theta_a$.

\subsubsection{Friction coefficient}

Before proceeding to the computation of the mean-free-path time, let us
check that the low-frequency friction coefficient for the STU black hole
that we can compute using the formula \eqref{mu_gen_geo} is consistent
with the result found in the literature \cite{Herzog:2007kh}.  In the
present case of the metric \eqref{STUmetricabc}, the formula
\eqref{mu_gen_geo} gives
\begin{align}
 \mu(\omega)&={2m\pi\ap\over \gamma(r_H)}+\CO(\omega).\label{mu_lowfreq_STU}
\end{align}
On the other hand, the drag force computed in \cite{Herzog:2007kh}
is\,\footnote{This is the drag force for the ``non-torque string'' of
\cite{Herzog:2007kh} which corresponds to no inflow of at the flavor
D-brane; see the discussion below \eqref{drag_ansatz}.  See Refs.\
\cite{Herzog:2006se, Caceres:2006dj, Herzog:2007kh} for the relation
between the strings with and without inflow.}
\begin{align}
\CF=-{\gamma(r_{ws})\over 2\pi\alpha'}v,\label{Herzogdragforce}
\end{align} 
where $v$ is the velocity of the quark and $r_{ws}$ is the solution to
$\alpha(r_{ws})-v^2 \gamma(r_{ws})=0$.
In the non-relativistic limit, $v\to 0$, the admittance read off from
\eqref{Herzogdragforce} should become the same as the low-frequency
result \eqref{mu_lowfreq_STU}.  Using the fact that $r_{ws}\to r_H$ and
$p=mv$ in the $v\to 0$ limit, it is easy to see that
\eqref{Herzogdragforce} indeed reproduces the admittance \eqref{mu_lowfreq_STU}.

\subsubsection{Mean-free-path time}

For the three cases (i)--(iii) described above, let us use the formula
\eqref{tmfp_final} and compute $t_{\text{mfp}}$.  Consider the
$n$-charge case ($n=1,2,3$). 
For the background configuration \eqref{STU_bg_soln}, the 10-dimensional
metric of the STU black hole \eqref{STU gen2} induces the following metric:
\begin{align}
  ds^2&=-fH^{-{n+1\over 2}}(1-f^{-1}H^{2}\CA_t^2)dt^2
 +H^{n-1\over 2}\left({dr^2\over f}+r^2(dX^I)^2\right),\\
 \CA_t&=
 \sqrt{mq\,}\left( {1\over r^2+q} - {1\over r_H^2+q} \right),\qquad
 H=1+{q\over r^2},
\end{align}
where $q=\kappa r_H^2$.  Here, in addition to the $t,r$ part, we kept
the $X^I$ part of the metric \eqref{STU_bg_soln} also, because we would
like to consider the transverse fluctuations along $X^I$ directions.
Comparing this metric with the general expression \eqref{gen
non-rotating metric}, we find
\begin{align}
  h_t=H^{-{n+1\over 2}}(1-f^{-1}H^{2}\CA_t^2),\qquad 
 h_r= H^{n-1\over 2}, \qquad
 G=r^2 H^{n-1\over 2}.\label{hthrG_STU}
\end{align}
Therefore, from \eqref{tmfp_final},
\begin{align}
 t_{\text{mfp}}
 \sim 
 {1\over T\log\bigl(\eta\sqrt{\lambda}\bigr)},
 \qquad\qquad
 \eta={L_H\over T\sqrt{H(r_H)}}
 ={L_H\over T\sqrt{1+\kappa}}.
\label{tmfp_STU}
\end{align}
The computation of $\eta$, particularly $L_H$ in it, is slightly
complicated. So, we delegate the details of the calculation to Appendix
\ref{app:eta_STU} and simply present the results. For 1-, 2-, and
3-charge cases, $\eta$ is given respectively by
\begin{align}
\label{eta_STU1}
 \eta&=
 {4\pi \over 2+\kappa}\exp\left\{
 -2(2+\kappa)\int_\infty^1 {d\rho\over \rho^2-1}\left[
 {1\over \sqrt{(\rho^2+1+\kappa)((1+\kappa)\rho^2+1)}}-
 {1\over 2+\kappa}\right]
 \right\}
 \\
\label{eta_STU2}
 \eta&=
 {2\pi \over \sqrt{1+\kappa}}\exp\left\{
 -4\sqrt{1+\kappa}\int_\infty^1 {d\rho\over \rho^2-1}\left[
 {1\over\sqrt{(\rho^2+1)(\rho^2+1+2\kappa)}}-{1\over 2\sqrt{1+\kappa}} 
 \right]
 \right\}
 \\
\label{eta_STU3}
 \eta&
  =
 {4\pi \over (1+\kappa)(2-\kappa)}\exp\Biggl\{
-2(1+\kappa)^{3/2}(2-\kappa)
 \int_\infty^1 {d\rho\over \rho^2-1}
 \notag\\
 &\qquad\qquad
 \times \Biggl[
 {\rho\over\sqrt{(\rho^2+1+\kappa-\kappa^2)(\rho^4+(1+3\kappa)\rho^2-\kappa^3)}}
 -{1\over (1+\kappa)^{3/2}(2-\kappa)}
 \Biggr]
 \Biggr\}
\end{align}
The small $\kappa$ expansion of $\eta$ is presented in
\eqref{eta_STU_expn1}--\eqref{eta_STU_expn3}.

\begin{figure}[tbp]
  \begin{center}
  \epsfxsize=8cm \epsfbox{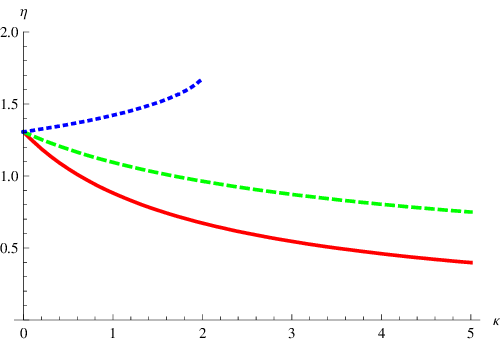} \caption{\sl Behavior of
   $\eta$ versus $\kappa$, for the 1-charge (solid red), 2-charge (dashed
   green), and 3-charge (dotted blue) cases.  The range of $\kappa$ is
   determined by the condition $T>0$ and the thermodynamical stability
   \eqref{thermo_stab} to be $\kappa<2$ for the 1- and 3-charge cases
   and $\kappa<1$ for the 2-charge case.}  \label{fig:eta_vs_kappa}
  \end{center}
\end{figure}
In Figure \ref{fig:eta_vs_kappa}, we have plotted the behavior of $\eta$
as we change $\kappa$.  Because $\eta$ appears in the denominator of the
expression for $t_{\text{mfp}}$, we observe the following: for the 1-
and 2-charge cases, $t_{\text{mfp}}$ gets longer as we increase the
chemical potential keeping $T$ fixed, while for the 3-charge case,
$t_{\text{mfp}}$ gets shorter as we increase the chemical potential
keeping $T$ fixed.

One may find it counter-intuitive that $t_{\text{mfp}}$ increases as we
increase chemical potential with $T$ fixed in the 1- and 2-charge cases,
based on the intuition that a larger chemical chemical potential means
higher charge density and thus more constituents to obstruct the motion
of the Brownian particle.  However, such intuition is not correct.  What
we know instead is that, if we increase the charge with the mass fixed,
then the entropy decreases, as one can see from the entropy formula for
charged black holes.  So, if we interpret entropy as the number of
``active'' degrees of freedom which can obstruct the motion of the
Brownian particle, then this suggests that $t_{\rm mfp}$ should increase
as we increase the charge with the mass fixed.  We did numerically check
that this is indeed true for all the 1-, 2- and 3-charge cases.

\section{Discussion}
\label{sec:conclusions}

We studied Brownian motion in the AdS/CFT setup and computed the time
scales characterizing the interaction between the Brownian particle and
the CFT plasma, such as the mean-free-path time $t_{\text{mfp}}$, by
relating them to the 2- and 4-point functions of random force.  We found
that there is an IR divergence in the computation of $t_{\text{mfp}}$
which we regularized by introducing an IR cutoff near the horizon.
Here let us discuss the issues involved in the procedure and the
implication of the result.

First, we note that the relation between $t_{\text{mfp}}$ and random
force correlators was derived using a simple classical model that we
proposed in subsection \ref{subsec:simple_model}.  As discussed in that
subsection, because the model is based on a kinetic theory picture that
is valid for weak coupling, its applicability to strongly coupled
plasmas is not obvious.  However, because of the simplicity of the
model, we believe that it captures the essential physics of the system
and gives a qualitatively correct value of $t_{\text{mfp}}$. This must
be kept in mind when interpreting the resulting expression for
$t_{\text{mfp}}$.

A natural question that arises about our result is: $t_{\text{mfp}}$ is
the mean-free-path time for what particle? First of all, one can wonder
whether this is really a mean-free-path time in the first place, because
the nontrivial 4-point function was obtained by expanding the
Nambu--Goto action to the next leading order, which is a relativistic
correction to the motion of the bulk string.  So, isn't this a
relativistic correction to the kinetic term in the Langevin equation,
not to the random force?  However, recall the ``cloud'' picture of the
Brownian particle mentioned before; the very massive quark we inserted
is dressed with a cloud of polarized plasma constituents. The position
of the quark corresponds to the boundary endpoint of the bulk string,
while the cloud degrees of freedom correspond to the fluctuation modes
of the bulk string.  So, we are incorporating relativistic corrections
to these cloud degrees freedom (fluctuations) but not to the quark which
gets very heavy in the large $m$ limit and thus remains
non-relativistic.

So, what is happening is the following.  First, the constituents of the
background plasma kick the cloud degrees of freedom randomly and,
consequently, those cloud degrees of freedom undergo random motion, to
which we have incorporated relativistic corrections. Then these cloud
degrees of freedom, in turn, kick the quark, which is recorded as the
random force $F$ felt by the quark.  $F$ is non-Gaussian, or has
a nontrivial 4-point function, because the cloud that is interacting with
the quark is
relativistic.  The quark's motion, which is what is observed in
experiments, is certainly governed by the non-Gaussian random $F$ and
the frequency of collision events is given by $1/t_{\text{mfp}}$.
However, it is worth emphasizing that this $t_{\text{mfp}}$ is not a
mean-free-path time for the plasma constituents
themselves.\footnote{Ref.\ \cite{Bhattacharyya:2007vs} estimates the
mean-free-path of the plasma constituents to be $l_{\text{mfp}}\sim 1/T$
from the parameters of the hydrodynamics that one can read off from the
bulk gravity.\label{ftnt:Shiraz}}

We focused on the fluctuations in the noncompact AdS directions.
However, for example, in the case of STU black holes, the full spacetime
is AdS$_5\times S^5$ and the string can fluctuate in the internal $S^5$
directions as well.  Let us denote the fluctuations of the string in the
internal directions by $Y$, while the fluctuations in the AdS directions
continue to be denoted by $X$.  One may wonder if the computations of
the random force correlators such as $\ev{\CO_X \CO_X \CO_X \CO_X}$ are
affected by the $Y$ fields.  Here, we denoted the force by $\CO_X$ to
remind ourselves that the force is an operator conjugate to the bulk
field $X$.  The $Y$ fields do contribute to such quantities, because the
Nambu--Goto action expanded up to quartic order involves terms of the
form $X^2Y^2$.  However, as long as we are interested in quantities with
all external lines being $O_X$, such as $\ev{\CO_X \CO_X \CO_X \CO_X}$,
they only make loop contributions, which are down by factors of $\ap$.
Therefore, our leading order results do not change.

In the present paper, we focused on the case where the plasma has no net
momentum.  More generally, one can consider the case where the plasma
carries net amount of momentum and insert a quark in it.  The Brownian
motion in such situations were studied in \cite{Giecold:2009cg,
CasalderreySolana:2009rm} (see also \cite{Giecold:2009wi}) in AdS/CFT
setups.  It is interesting to generalize our computation of
$t_{\text{tmp}}$ to such cases.  Note the following, however: in
general, in the presence of a net background momentum, the string will
``trail back'' because it is pushed by the flow.  Unless one applies an
external force, the string will start to move and ultimately attain the
same velocity as the background plasma.  This final state is simply a
boost of the static situation studied in the present paper.  So, the
result of the current paper applies to this last situation too (after
rescaling due to Lorentz contraction).

The resulting expression for the mean-free-path time, {\it e.g.}\
\eqref{tmfp_ft0}, is quite interesting because of the logarithm.  As
mentioned around \eqref{tcoll_ft0}, this means that the Brownian
particle is experiencing $\sim\log\lambda$ collision events at the same
time.  Because $\lambda\propto N$, this is reminiscent of the fast
scrambler proposal \cite{Hayden:2007cs, Sekino:2008he} which claims
that, in theories that have gravity dual, $\sim\log N$ degrees of freedom are in
interaction with each other simultaneously.

In our previous paper \cite{deBoer:2008gu}, we claimed that
$t_{\text{mfp}}\sim 1/T$ based on dimensional analysis, but
\eqref{tmfp_ft0} says that there is an extra factor which cannot be
deduced on dimensional grounds.  Of course, we have to note the fact
that $t_{\text{mfp}}$ we computed in the present paper is not the time
scale of the constituents but of the Brownian particle (see also
footnote \ref{ftnt:Shiraz}).  In our previous paper
\cite{deBoer:2008gu}, we had $t_{\rm relax}^{\rm old}\sim
m/(T^2\sqrt{\lambda})$, $t_{\rm mfp}^{\rm old}\sim 1/(T\sqrt{\lambda})$
instead, which were nice because if we set $m\to T$ in $t_{\rm
relax}^{\rm old}$ we get $t_{\rm mfp}^{\rm old}$.  In the
\eqref{tmfp_ft0}, this is no longer the case, but now the relation
between $t_{\rm relax}$ and $t_{\rm mfp}$ is not so simple as we can see
from the fact that there is a nontrivial $\lambda$ dependence in the
impact per collision, $P(\omega=0)$ (Eq.\ \eqref{tmfp_ft0}).  It would
be interesting to find an improved microscopic toy model which  can relate
$t_{\rm relax}$ and $t_{\rm mfp}$.

Probably the most controversial issue in our computations is the IR
cutoff.  When regulating integrals such as \eqref{R4_IRdiv}, we cut off
the geometry at a proper distance $s\sim l_s$ away from the horizon, assuming
that the contribution from the rest of the integral is of the same
order.  This seems physically reasonable, but we do not have a proof.
One could also have tried to
put a cutoff at the point where the backreaction of the fundamental
string on the black hole geometry becomes important.  Since the
interaction of the string with the background is suppressed by
additional powers of the string coupling constant, the resulting cutoff
is presumably closer to the Planck length than the string length.

One might wonder whether the divergences disappear if we use a different
basis for correlation functions, such as the $r$-$a$ basis mentioned in
subsection \ref{ss:comment_basis}.  However, one can show that, after
adding appropriate counter terms near the boundary, the $r$-$a$ basis
correlators are all UV finite but are still IR divergent.\footnote{More
specifically, the divergence of the 4-point functions in the $r$-$a$
basis depends only on the number of $r$, $a$ indices and
$G_{rrrr},G_{rrra},G_{rraa},G_{raaa}$ are all IR divergent although the
degree of divergence becomes lower in the this order, while $G_{aaaa}$
vanishes.  Note that there $r$-$a$ correlators are related to one
another by the fluctuation-dissipation relations (see {\it e.g.},
\cite{Carrington:2006xj}).  }  This again suggests that the IR
divergence is not an artifact but of a physical origin.

Related to the above statements, it is interesting to note that the
mean-free-path at weak coupling \cite{Arnold:1997gh}
\begin{align}
 \lambda_{\rm mfp,weak}\sim{1\over g^4_{\rm YM}T\ln(1/g^2_{\rm YM})}
 \label{t_mfp_weak}
\end{align}
has a form tantalizingly similar to \eqref{tmfp_ft0_0}. In particular,
the log in \eqref{t_mfp_weak} is coming from an IR divergence cut off by
non-perturbative magnetic effects \cite{Arnold:1997gh}, while the log in
\eqref{tmfp_ft0_0} was also coming from an IR divergence that we regularized
by introducing an IR cutoff.  It would be interesting to study whether
there is a relation between the weakly and strongly coupled descriptions
of the IR divergences and the physical interpretation of the IR cutoffs.

\section*{Acknowledgments}

We would like to thank J.~Casalderrey-Solana, C.~P.~Herzog, V.~Hubeny,
E.~Laenen, K.~Papadodimas, M.~Rangamani, K.~Skenderis, S.~Sugimoto,
T.~Takayanagi, D.~Teaney, and E.~Verlinde for valuable discussions.  We also thank
K. Schalm and B.~van Rees for collaboration in the early stage of the
project.
M.S. would like to thank the organizers of the workshops ``Tenth
Workshop on Non-Perturbative Quantum Chromodynamics'' at l'Institut
Astrophysique de Paris for stimulating environments, and IPMU and Yukawa
Institute for Theoretical Physics where part of the current work was
done, for hospitality.
The work of M.S. was supported by an NWO Spinoza grant. The work of
J.d.B. and M.S. is partially supported by the FOM foundation. The work
of A.N.A. was supported in part by a VIDI Innovative Research Incentive
Grant from NWO.

\appendix

\section{Normalizing solutions to the wave equation}
\label{sec:normalizing}

As explained in subsection \ref{subsec:bulk_BM} or more generally in
subsection \ref{subsec:BM_bulk_gen's}, the normalized modes
$\{u_\omega\}$ are proportional to $f_\omega$ of the form
\eqref{f_ito_g}; namely, $u_\omega(t,r)\propto e^{-i\omega
t}f_\omega(r)$.  Here, we fix the normalization and derive the expansion
\eqref{Xexpn_neutral} or more generally \eqref{Xexpn_gen}.

The analogue of the Klein--Gordon inner product for functions
$f(t,r),g(t,r)$ satisfying the equation of motion \eqref{gen eqm} is
\cite{deBoer:2008gu}
\begin{align}
 (f,g)_\Sigma=-{i\over 2\pi\ap}\int_\Sigma \sqrt{\widetilde g\,}\,
 n^\mu G\, (f\partial_\mu g^*-\partial_\mu f\, g^*),
\label{inner_prod}
\end{align}
where $\Sigma$ is a Cauchy surface in the $t,r$ part of the metric
\eqref{gen non-rotating metric}. $\widetilde g$ is the induced metric on
$\Sigma$ and $n^\mu$ is the future-pointing unit normal to $\Sigma$.

We want to normalize $f_\omega$ using this norm \eqref{inner_prod}.  In
the present case, there is the following simplification to this
procedure.  Near the horizon $r\sim r_H$, the action \eqref{Gen NG
expansion} reduces to
\begin{align}
 S_0&\approx {G(r_H)\over 4\pi\ap}\int dt\, dr_*
 \left[(\partial_t X)^2-(\partial_{r_*}X)^2\right].\label{action_NH}
\end{align}
Therefore, in this region and in the tortoise coordinate system, $X$ is
just like a massless Klein--Gordon scalar in flat space.
Correspondingly, the contribution to the norm \eqref{inner_prod} from
the near horizon region is
\begin{align}
 -{iG(r_H) \over 2\pi\ap}\int_{r_*\sim-\infty} dr_*
 (f\, \partial_t g^*-\partial_t f\, g^*),\label{inner_prod_NH}
\end{align}
where as $\Sigma$ we took the constant $t$ surface.  This is the usual
Klein--Gordon inner product for the theory \eqref{action_NH}, up to
overall normalization.
Of course, there is  a contribution to the inner product from regions away
from the horizon. However, because the near-horizon region is
semi-infinite in the tortoise coordinate $r_*$ (recall that $r=r_H$
corresponds to $r_*=-\infty$), the normalization of solutions is
completely determined by this region where the inner product is simply
\eqref{inner_prod_NH}. This means that the canonically normalized mode
expansion is given by
\begin{align}
 X(t,r)=\sqrt{2\pi\ap\over G(r_H)}\int_0^\infty {d\omega\over 2\pi}
 {1\over \sqrt{2\omega}}
 \left[ f_\omega(r)e^{-i\omega t}a_\omega+ f_\omega(r)^* e^{i\omega t}a_\omega^\dagger\right].\label{X(t,r)_expn_free}
\end{align}
where $f_\omega(r)$ behaves near the horizon as
\begin{align}
 f_\omega(r)\to e^{i\omega r_*} +e^{i\theta_\omega}e^{-i\omega r_*},\qquad
 r\to r_H\quad (r_*\to-\infty)\label{f_NH_asympt_app}
\end{align}
with some $\theta_\omega\in\bbR$.
If we can find such $f_\omega(r)$, then $a,a^\dagger$ satisfy the
canonically normalized commutation relation \eqref{CCR_a}.

\section{Low energy solutions to the wave equation}
\label{sec:lowE_wave}

Here, we study the solution to the wave equation \eqref{EOMgen}, or more
generally \eqref{gen eqm}, satisfying an appropriate boundary condition
(the Neumann boundary condition \eqref{Neumanbc_rc} or normalizable
boundary condition at infinity), for very small frequencies $\omega$.
We see that the solutions become trivial plane waves in the limit.

The general wave equation \eqref{gen eqm} can be written in the
frequency space as
\begin{equation}
\label{gen_eqm_app}
 \left[\omega^2+\sqrt{\frac{h_t}{h_r}}\,\frac{f}{G}\,
 \partial_r\!\left(\sqrt{\frac{h_t}{h_r}}\,f G \partial_r\right)\right]
 X_\omega(r)=0.
\end{equation}
Very close to the horizon, this becomes
\begin{align}
 \left[\omega^2+16\pi^2 T^2(r-r_H)\partial_r\bigl((r-r_H)\partial_r\bigr)\right]
 X_\omega(r)=0.
\end{align}
This means that the linearly independent solutions are
\begin{align}
 g_{\pm \omega}
 =
 \exp\left[\pm i{\omega\over 4\pi T}\log\left({r-r_H\over L_1}\right)\right]
 \label{NH_sol_waveeq}
\end{align}
where $L_1$ is a length scale which is arbitrary at this point.  The
$\pm$ signs here correspond to outgoing and ingoing waves.
We are considering the small $\omega$ limit but, no matter how small
$\omega$ is, we can always consider a region very close to the horizon
so that ${\omega\over 4\pi T}\log({r-r_H\over L_1})=\CO(1)$, namely
${r-r_H\over L_1}\lesssim e^{-4\pi T/\omega}$.  In such a region, we
cannot expand the exponential and should keep the full exponential
expression  \eqref{NH_sol_waveeq}.  In other words, the precise limit we
are taking is
\begin{align}
 \omega\to 0,\qquad {\omega\over T}\log\Bigl({r-r_H\over L_1}\Bigr):~\text{fixed.}\label{preciselim}
\end{align}

Now, consider the region not so close to the horizon.  For small
$\omega$, we can ignore the $\omega^2$ term in \eqref{gen_eqm_app}, obtaining
\begin{align}
 X_\omega&=B_1+B_2\int_\infty^r {dr'\over f(r')G(r')}\sqrt{h_r(r')\over h_t(r')}
 +\CO(\omega^2),\label{gen_sol_waveeq_larger}
\end{align}
where $B_1,B_2$ are constant.  For $r\approx r_H$, this gives
\begin{align}
 X_\omega&= B_1+{B_2\over 4\pi TG(r_H)}\log\left({r-r_H\over L_H}\right)+\CO(r-r_H) 
 \qquad\qquad (r\sim r_H).\label{NH_sol_waveeq2}
\end{align}
Here, we defined the constant $L_H$ by 
\begin{align}
 \int_\infty^r {dr\over fG}\sqrt{h_r\over h_t}
 ={1\over 4\pi T G(r_H)}\log\left({r-r_H\over L_H}\right)+\CO(r-r_H).\label{def_LH_app}
\end{align}
Because it will turn out to be convenient to choose $L_H=L_1$, we will
set $L_H=L_1$ henceforth.  On the other hand, for large $r$,
\eqref{gen_sol_waveeq_larger} gives (assuming the large $r$ behavior
\eqref{asymptcond_h}, \eqref{asymptcond_G} of functions $h_t,h_r,G$),
\begin{align}
 X_\omega &=B_1-{B_2\over 3r^3}+\CO(\omega^2).\label{largeXomega}
\end{align}

We can determine $B_1,B_2$ by comparing these small-frequency solutions
between the very-near-horizon region and the not-so-near-horizon region.
For small frequencies $\omega$, \eqref{NH_sol_waveeq} becomes
\begin{align}
 X_\omega\approx 
 1\pm i{\omega\over 4\pi T}\log\left({r-r_H\over L_H}\right)+\CO(\omega^2).
\end{align}
Comparing this with \eqref{NH_sol_waveeq2}, we determine 
\begin{align}
 B_1=1+\CO(\omega^2),\qquad
 B_2=\pm i\omega G(r_H)+\CO(\omega^2).\label{B1B2}
\end{align}
Therefore, the linearly independent (outgoing/ingoing) solutions are
\begin{align}
 g_{\pm \omega}(r)
 =
 \begin{cases}
  \displaystyle
  \exp\left[\pm i{\omega\over 4\pi T}\log\left({r-r_H\over L_H}\right)\right]
  &\qquad  r\sim r_H
  \\[3ex]
  \displaystyle
  1
  \pm{i\omega G(r_H)\over 3r^3}
  &\qquad  r\gg r_H
 \end{cases}
 \label{Xpm_asympt}
\end{align}

The general solution $X_\omega$ is given by the linear combination of
the outgoing and ingoing solutions $g_{\pm \omega}$.
If we want to construct a normalizable solution that
vanishes as $r\to \infty$ then, from the $r\gg r_H$ behavior of
\eqref{Xpm_asympt}, the linear combination to take is
\begin{align}
X_\omega^{\text{(norm)}}
 &=
g_\omega- g_{-\omega}.
\end{align}
If we did not take $L_1=L_H$, the two terms would be multiplied by
$\exp[\mp i{\omega\over 4\pi T}\log({L_H\over L_1})]$ respectively.  Note
that our expressions are correct up to $\CO(\omega^2)$ terms.  The
near-horizon behavior of this is
\begin{align}
X_\omega^{\text{(norm)}}
 \approx
  \exp\left[  i{\omega\over 4\pi T}\log\left({r-r_H\over L_H}\right)\right]
 -\exp\left[- i{\omega\over 4\pi T}\log\left({r-r_H\over L_H}\right)\right].\label{Xnorm_NH}
\end{align}
Therefore, if define the tortoise coordinate $r_*$ to have the following
behavior the horizon:
\begin{align}
r_*\approx {1\over 4\pi T}\log\left({r-r_H\over L_H}\right)\label{rs_def_LH_app}
\end{align}
then the near-horizon behavior \eqref{Xnorm_NH} simply becomes 
\begin{align}
X_\omega^{\text{(norm)}}
 \approx e^{i\omega r_*}-e^{-i\omega r_*}\qquad (r\approx r_H).\label{NH_wave_normble}
\end{align}
Let us elaborate on this point slightly more.  In the near horizon
region, in general we can have
\begin{align}
X_\omega^{\text{(norm)}}
 \approx e^{i\omega r_*}-e^{i\alpha_\omega}e^{-i\omega r_*}.
 \label{alpha_omega}
\end{align}
where $\alpha_\omega$ is some phase. The fact that
\eqref{NH_wave_normble} is correct up to $\CO(\omega)$ means is that, if
we take $r_*$ to be given by \eqref{rs_def_LH_app}, then
$\alpha_\omega=\CO(\omega^2)$ as $\omega\to 0$. In particular, unless we
choose $L_H$ to be the one given by \eqref{def_LH_app}, the $\omega\to
0$ behavior of $\alpha_\omega$ will contain an $\CO(\omega)$ term.

Next, let us consider imposing a Neumann boundary condition $\partial_r
X=0$ at $r=r_c\gg r_H$ instead.  Set the general solution to be
\begin{align}
 X_\omega=g_{\omega}+ C g_{-\omega}.
\end{align}
then the Neumann boundary condition $X_\omega'(r_c)=0$ gives
\begin{align}
 C&=-{g_\omega'(r_c)\over g_{-\omega}'(r_c)}
 =-{
 -{i\omega G(r_H)\over r_c^4}+\CO(\omega^2)
 \over
 {i\omega G(r_H)\over r_c^4}+\CO(\omega^2)}
 =1+\CO(\omega),\label{C-overC+}
\end{align}
where we used the second equation in \eqref{Xpm_asympt}.  Comparing this
result with \eqref{g_NH_asympt} and \eqref{f_NH_asympt}, we find that
the modes $f_\omega$ satisfying the Neumann boundary condition are given
by, at low frequencies,
\begin{align}
\begin{split}
  g_\omega(r)&=e^{i\omega r_*},\qquad
 \theta_\omega=0,\qquad
 f_\omega(r)=e^{i\omega r_*}+e^{-i\omega r_*}.
\end{split}
\label{lowfreq_f,g}
\end{align}
This is consistent with the explicit result for AdS$_3$ in
\eqref{fpm_def}, \eqref{BoverA}.  
So, for very small $\omega$, the solution $f_\omega(r)$ is a simple sum
of outgoing and ingoing waves, which are just plane waves.\footnote{For related
observations on the triviality of the solution in the low frequency
limit, see \cite{Iqbal:2008by}.}
Because $g_{\pm}(r)\to 1$ as $r\to
\infty$, we have
\begin{align}
 f_\omega(r=r_c)\approx 2.\label{f(r=rc)=2}
\end{align}
Because of the $\CO(\omega)$ ambiguity in \eqref{C-overC+},
$\theta_\omega=\CO(\omega)$ as $\omega\to 0$ (cf.\ comments below
\eqref{alpha_omega}).

In the tortoise coordinate, the limit \eqref{preciselim} we are taking
can be written as
\begin{align}
 \omega\to 0, \qquad
 \beta,~\omega r_*:~\text{fixed}.
 \label{low_freq_lim_app}
\end{align}

\section{Various propagators and their low frequency limit}
\label{sec:lowE_propagators}

The quadratic action for a string embedded in the AdS$_d$ black hole
spacetime
\begin{align}
 ds^2&=-{h_t(r)f(r)}dt^2+{h_r(r)\over f(r)}dr^2+G(r) dX^2,
\end{align}
(Eq.\ \eqref{gen non-rotating metric}) is given by
\begin{gather}
 S_0
 ={1\over 4\pi \ap}\int dt\,dr
 \left[\sqrt{h_r\over h_t}\,{G\over f}(\partial_t X)^2
 -\sqrt{h_t\over h_r}\,Gf\,(\partial_r X)^2\right].\label{Squad_app}
\end{gather}
We would like to regard this system as a thermal field theory at
temperature $T$, and derive the relation among various propagators
(Green functions) and the solutions to the wave equation.  We present
the result for the general metric \eqref{gen non-rotating metric}, but
if one wants the results for the simpler neutral case \eqref{AdSdBH},
set $f=r^2h$, $h_t=h_r=1$, $G=r^2$.

Let us define Wightman, Feynman, retarded, and advanced propagators as
\begin{align}
\begin{split}
  D_W(t-t',r,r')    &=\ev{X(t,r)X(t',r')},\\
 D_F(t-t',r,r')    &=\ev{\CT[X(t,r)X(t',r')]},\\
 D_{\rm ret}(t-t',r,r')&=\theta(t-t')\ev{[X(t,r),X(t',r')]},\\
 D_{\rm adv}(t-t',r,r')&=-\theta(t'-t)\ev{[X(t,r),X(t',r')]}.
\end{split}
\end{align}
We impose a Neumann boundary condition for $X(r,t)$ at $r=r_c$, so the
propagators satisfy the same Neumann boundary condition. Using the wave
equation
\begin{gather}
 \left[-{G\over h_t f}\partial_t^2
 +{1\over \sqrt{h_t h_r}}\partial_r\left(\!\sqrt{h_t\over h_r}\, Gf\,\partial_r\right)\right]\! X=0
\end{gather}
and the canonical commutation relation
\begin{align}
 [X(t,r),\partial_t X(t,r')]&={2\pi i\ap }\sqrt{h_t\over h_r}\, {f\over G}\,\delta(r-r'),
\end{align}
we can show that these propagators satisfy
\begin{align}
 \left[-{G\over h_t f}\partial_t^2
 +{1\over \sqrt{h_t h_r}}\partial_r\left(\!\sqrt{h_t\over h_r}\, Gf\,\partial_r\right)\right]\!
 D_{W}(t-t',r,r')
 &=0,\\
 \left[-{G\over h_t f}\partial_t^2
 +{1\over \sqrt{h_t h_r}}\partial_r\left(\!\sqrt{h_t\over h_r}\, Gf\,\partial_r\right)\right]\!
 D_{\text{F,ret,adv}}(t-t',r,r')
 &={2\pi i \ap \over \sqrt{-g}}\delta(t-t')\delta(r-r').\label{waveeq_D_F,ret,adv}
\end{align}
where $\sqrt{-g}=\sqrt{h_t h_r}$.

As in \eqref{X(t,r)_expn_free}, the field $X$ can be expanded as
\begin{align}
 X(t,r)={\sqrt{2\pi\ap\over G(r_H)}}\int_0^\infty {d\omega\over 2\pi}
 {1\over \sqrt{2\omega}}
 \left[ f_\omega(r)e^{-i\omega t}a_\omega
 + f_\omega(r)^* e^{i\omega t}a_\omega^\dagger\right],
\end{align}
where
\begin{align}
 f_\omega(r)&=g_\omega(r)+e^{i\theta_\omega}g_{-\omega}(r)\label{f_ito_g_app}
\end{align}
and $g_\omega(r)$ behaves near the horizon as
\begin{align}
 g_\omega(r)&\approx e^{i\omega r_*} \qquad (r\sim r_H).\label{g_asympt_NH_app}
\end{align}
The phase $\theta_\omega$ is determined by the Neumann boundary
condition at $r=r_c$ that $f_\omega$ satisfies.  Since the system is at
temperature $T$, the expectation value of $a,a^\dagger$ is given by
\eqref{rel_x_aadagger}.  It is then easy to show that the Wightman
propagator can be written as
\begin{align}
 D_W(\omega,r,r')&=
 {2\pi\ap\over G(r_H)}
 {f_{\omega}(r)f_{-\omega}(r')\over 2\omega(1-e^{-\beta\omega})},\label{D_W_ito_f}
\end{align}
where $f_{-\omega}=f_{\omega}^*$.

We would like to express other propagators $D_{\rm adv,ret,F}$ in terms of
$f_\omega,g_\omega$.   Note that
\begin{align}
 D_F(\omega,r,r')
 &=D_W(\omega,r,r')+D_{\text{adv}}(\omega,r,r')
 =D_W(-\omega,r',r)+D_{\text{ret}}(\omega,r,r').\label{rel_Ds}
\end{align}
Because we have already obtained $D_W$ in \eqref{D_W_ito_f}, if we know
one of $D_F$, $D_{\rm ret}$, and $D_{\rm adv}$, we can obtain all other
propagators.  Here, let us consider $D_{\rm adv}$.  From
\eqref{waveeq_D_F,ret,adv}, $D_{\rm adv}(\omega,r,r')$ should satisfy
\begin{align}
 \left[{G\over h_t f}\omega^2
 +{1\over \sqrt{h_t h_r}}\partial_r\left(\!\sqrt{h_t\over h_r}\, Gf\,\partial_r\right)\right]\!
 D_{\text{adv}}(\omega,r,r')
 &={2\pi i \ap\over\sqrt{-g}}\delta(r-r').\label{Dadv_ansatz_app}
\end{align}
If $r\neq r'$, this is the same as the wave equation that
$f_{\omega},g_\omega$ satisfy.  Therefore, take the ansatz
\begin{align}
 D_{\text{adv}}(\omega,r,r')=A[\theta(r-r')g_\omega(r')f_\omega(r)+\theta(r'-r)g_\omega(r)f_\omega(r')].
\end{align} 
This satisfies the correct boundary condition (Neumann) at $r,r'=r_c$
and furthermore satisfies the purely outgoing boundary condition at the
horizon, which is appropriate for an advanced correlator.
Using the fact that both $f,g$ satisfy the wave equation, we find
\begin{align}
 \left[{G\over h_t f}\omega^2
 +{1\over \sqrt{h_t h_r}}\partial_r\left(\!\sqrt{h_t\over h_r}\, Gf\,\partial_r\right)\right]\!
D_{\rm adv}
 ={A\delta(r-r')\over\sqrt{-g}}\, \sqrt{h_t\over h_r}\, Gf\,(g_\omega\partial_r f_\omega-\partial_r g_\omega\,f_\omega).
\end{align}
Therefore,
\begin{align}
 A={2\pi i\ap}\sqrt{h_r\over h_t}\, {1\over Gf}\, {1\over g_\omega \partial_r f_\omega -\partial_r g_\omega \,f_\omega}
  ={2\pi i\ap  \over G(r_H)\,(g_\omega \partial_{r_*} f_\omega-\partial_{r_*} g_\omega\,f_\omega)}.\label{valueA}
\end{align}
Using the wave equation for $f_\omega,g_\omega$, it is easy to show that this
expression does not depend on $r$.  By taking $r\to r_H$ and using
\eqref{f_ito_g_app}, \eqref{g_asympt_NH_app}, 
\begin{align}
 A& =-{\pi \ap e^{-i\theta_\omega}\over G(r_H)\, \omega}.
\end{align}
So, the advanced propagator is given by
\begin{align}
 D_{\rm adv}(\omega,r,r')
 &=-{\pi \ap e^{-i\theta_\omega}\over G(r_H)\, \omega}
 \bigl[\theta(r-r')g_\omega(r')f_\omega(r)+\theta(r'-r)g_\omega(r)f_\omega(r')\bigr].\label{Dadv_expl}
\end{align} 

In the low frequency limit, the expressions for the propagators
simplify, as we saw in Appendix \ref{sec:lowE_wave}.  The precise limit
we are considering is \eqref{low_freq_lim_app}.  First, the Wightman
propagator \eqref{D_W_ito_f} becomes, because of \eqref{lowfreq_f,g},
\begin{align}
 D_W(\omega,r,r')&=
 {\pi\ap\over G(r_H)}
 {(e^{i\omega r_*}+e^{-i\omega r_*})(e^{i\omega r_*'}+e^{-i\omega r_*'})
 \over \omega(1-e^{-\beta\omega})}
 \qquad (\text{small $\omega$}).
 \label{D_W-lowE}
\end{align}
Similarly, the advanced propagator \eqref{Dadv_expl} becomes
\begin{align}
 D_{adv}(\omega,r,r')
 &=-{\pi \ap \over G(r_H)\, \omega}
 \left[\theta(r_*-r_*')e^{i\omega r_*'}(e^{i\omega r_*}+e^{-i\omega r_*})
 +\theta(r_*'-r_*)e^{i\omega r_*}(e^{i\omega r_*'}+e^{-i\omega r_*'})\right]
 \notag\\
 &=-{\pi \ap \over G(r_H)}
 {e^{i\omega (r_*+r_*')}+e^{-i\omega |r_*-r_*'|}\over \omega}
 \qquad (\text{small $\omega$}).
 \label{D_adv-lowE}
\end{align} 
Using  the relation \eqref{rel_Ds}, the Feynman propagator is
\begin{align}
 D_F(\omega,r,r')&=
 {\pi\ap\over G(r_H)}\left[
 {(e^{i\omega r_*}+e^{-i\omega r_*})(e^{i\omega r_*'}+e^{-i\omega r_*'})
 \over \omega(1-e^{-\beta\omega})}
 - {e^{i\omega (r_*+r_*')}+e^{-i\omega |r_*-r_*'|}\over \omega}
 \right]
 \qquad (\text{small $\omega$}).
 \label{D_F-lowE}
\end{align}

In particular, consider the case where one of the points is at the UV
cutoff, $r'=r_c$.  From \eqref{f(r=rc)=2}, we have
\begin{align}
\begin{split}
    D_F(\omega,r,r_c)&=
 {2\pi\ap\over G(r_H)}\left[
 {e^{i\omega r_*}+e^{-i\omega r_*}
 \over \omega(1-e^{-\beta\omega})}
 - {e^{i\omega r_*}\over \omega}
 \right],\\
 D_W(\omega,r,r_c)&=
 {2\pi\ap\over G(r_H)}
 {e^{i\omega r_*}+e^{-i\omega r_*}
 \over \omega(1-e^{-\beta\omega})}.
\end{split} 
\label{lowE_D_F,W_app}
\end{align}

\section{Holographic renormalization and Lorentzian AdS/CFT}
\label{app:holo_ren}

In this Appendix, we discuss how to compute correlation function using
the AdS/CFT dictionary for the total force $F$ which is dual to the
worldsheet field $X$.  As we explained in subsection
\ref{subsec:holo_ren}, this involves holographic renormalization
(see {\it e.g.}\ \cite{Skenderis:2002wp}) of the worldsheet action.  Furthermore, if we
want to compute real time correlation functions in a black hole (finite
temperature) geometry, we should apply the rules of Lorentzian AdS/CFT
\cite{Skenderis:2008dg, vanRees:2009rw}.

\subsection{Holographic renormalization}
\label{subsec:app:holo_ren}

First, let us consider the holographic renormalization of the worldsheet
action.  For this, only the asymptotic behavior of the action near the
boundary is relevant.  Therefore, as the background geometry, we can
consider the Poincar\'e AdS geometry obtained by setting $T=0$
\eqref{AdSdBH}:
\begin{align}
 ds^2&=-r dt^2+{dr^2\over r^2}+r^2(dX^I)^2.
\label{T=0AdS_app}
\end{align}
for which the worldsheet action becomes
\begin{align}
 S_{\text{bare}}&=S_0+S_{\text{int}},\\
 S_{0}&={1\over 4\pi\ap}\int_\Sigma dt\,dr ({\dot X^2}-{r^4}X'^2),\qquad
 S_{\text{int}}={1\over 16\pi\ap}\int_\Sigma dt\,dr ({\dot X^2}-{r^4}X'^2 )^2.
 \label{Sbare_app-1}
\end{align}
Here, we considered only one of the polarizations, say $X^1$, and denoted
it by $X$.  $\Sigma$ is the worldsheet,
\begin{align}
 \Sigma=\{(t,r)\,|\,t\in\bbR,\, 0\le r\le r_c\},
\end{align}
and $\dot{}=\partial_t$, ${}'=\partial_r$.  $r_c$ is a UV cutoff.  For
computational convenience, let us we rescale $X\to \sqrt{2\pi\ap}\,X$
and set $\kappa=\pi\ap$, so that
\begin{align}
 S_{0}&={1\over 2}\int_\Sigma dt\,dr ({\dot X^2}-{r^4}X'^2),\qquad
 S_{\text{int}}={\kappa\over 4}\int_\Sigma dt\,dr ({\dot X^2}-{r^4}X'^2 )^2.
 \label{Sbare_app-2}
\end{align}
The equation of motion is
\begin{align}
 -\partial_t^2 X+\partial_r (r^4\partial_r X)
 =\kappa[-\partial_t(H\partial_t X)+\partial_r (H r^4\partial_r X)],\qquad
 H\equiv -\dot X^2+r^4X'^2.
 \label{eom_app}
\end{align}

Let us solve the equation of motion \eqref{eom_app} by expanding
$X(t,r)$ in the coupling $\kappa$ as
\begin{align}
 X(t,r)=Y(t,r)+\kappa Z(t,r)+\CO(\kappa^2)
 \label{FGexpn_app}
\end{align}
and furthermore expanding $Y,Z$ around $r=\infty$ as
\begin{align}
\begin{split}
 Y(t,r)&=y_{(0)}(t)+{y_{(1)}(t)\over r}+{y_{(2)}(t)\over r^2}+{y_{(3)}(t)\over r^3}+\cdots,\\
 Z(t,r)&=z_{(0)}(t)+{z_{(1)}(t)\over r}+{z_{(2)}(t)\over r^2}+{z_{(3)}(t)\over r^3}+\cdots.
\end{split} 
\label{FGexpn_YZ_app}
\end{align}
Henceforth, we will ignore quantities of $\CO(\kappa^2)$.
The expansion for $X$ itself is
\begin{align}
 X(t,r)&=x_{(0)}(t)+{x_{(1)}(t)\over r}+{x_{(2)}(t)\over r^2}+{x_{(3)}(t)\over r^3}+\cdots,
 \qquad x_{(i)}=y_{(i)}+\kappa z_{(i)}.
\label{FGexpn_X_app}
\end{align}
By substituting this expansion into \eqref{eom_app} and comparing
coefficients, one readily finds that the following is a solution:
\begin{subequations}
\begin{align}
  y_{(0)}&=\text{any}\equiv J,\qquad
 y_{(1)}=0,\qquad
 y_{(2)}=-{1\over 2}\ddot J,\qquad
 y_{(3)}=\text{any},
\label{FGexpn_Y_sol_app}
 \\
 z_{(0)}&=0,\qquad
 z_{(1)}=0,\qquad
 z_{(2)}=-\dot{J}^2\ddot{J},\qquad
 z_{(3)}=\text{any}.
\label{FGexpn_Z_sol_app}
\end{align}
\label{FGexpn_Y,Z_sol_app}
\end{subequations}
The expression for $X$ is
\begin{align}
 x_{(0)}=J,\qquad
 x_{(1)}=0,\qquad x_{(2)}=-{1\over 2}\ddot J -\kappa \dot{J}^2\ddot{J}
 ,\qquad
 x_{(3)}=\text{any}.
\label{FGexpn_X_sol_app}
\end{align}
Note that $X(r,t)\to J(t)$ as $r\to\infty$; namely, $J$ is the
non-normalizable mode which can be thought of as a source for the dual
operator $\CO_X=F$ on the boundary.  On the other hand, $x_{(3)}$ is the
normalizable mode which roughly corresponds to the expectation value of
the operator $F$\@. We will make this latter statement more precise below.

If we plug the solution \eqref{FGexpn_X_sol_app} into the action 
\eqref{Sbare_app-2}, we obtain the following on-shell action:
\begin{align}
 S_{\text{bare,on-shell}}&={\kappa\over 2}\int_\Sigma d^2x\, J\dot{J}^2 \ddot{J}
 +\int_{\partial\Sigma} dt\, \left[
\left( -{1\over 2} r J\ddot{J}
 -\kappa r J \dot J^2 \ddot J
 \right)
 -{\kappa\over 4} r J\dot{J}^2\ddot J
 \right]\notag\\
 &\sim \int_{\partial\Sigma} dt\, \left[
 -{1\over 2} r J\ddot{J}
 -{3\kappa\over 4} r J\dot{J}^2\ddot J
 \right]+\text{(finite)}\notag\\
 &\sim \int_{\partial\Sigma} dt\, \left[
 {1\over 2} r \dot{J}^2
 +{\kappa\over 4} \dot{J}^4
 \right]+\text{(finite)}.
\end{align}
In going to the second line we performed the $r$ integration, and in
going to the last line we integrated by parts. This is divergent, but
the divergence can be canceled by introducing the following counter
terms:
\begin{align}
 S_{\text{ct}}
 &= \int_{\partial\Sigma} dt\, \sqrt{-\gamma}\left[
 {1\over 2} r^2  (\nabla_\gamma X)^2
 -{\kappa\over 4} \bigl(r^2 (\nabla_\gamma X)^2\bigr)^2
 \right],
\label{Sct_app}
\end{align}
where $\Box_\gamma=-{1\over r^2}\partial_t^2$ is the Laplacian for the
metric $\gamma$ induced on the boundary $r=r_c$.  Likewise,
$(\nabla_\gamma X)^2=-{1\over r^2}(\partial_t X)^2$.  If we define the
metric $\gamma'$ induced on the boundary of the worldsheet at $r$, then
$\gamma'_{tt}=-r^2(1-\dot X^2)$ and $\int dt\sqrt{-\gamma'_{tt}}$
reproduces \eqref{Sct_app} (also recall that we have rescaled
$X\to\sqrt{2\pi\ap}\,X$).

To remove the divergence from the ``bare'' action \eqref{Sbare_app-2},
we take $S_{\text{ren}}=S_{\text{bare}}+S_{\text{ct}}$ as our total
action.  The on-shell variation of this total action evaluates to
\begin{align}
 \delta S_{\text{ren,on-shell}}&=\int_{\partial\Sigma} dt\sqrt{-\gamma}\biggl(
 -r^2(\partial_n X+\Box_\gamma X)+\kappa r^4 \left[
 (\nabla X)^2 \partial_n X+3 (\nabla_\gamma X)^2 \Box_\gamma X
 \right]
 \biggr)\delta X,\label{dSren,os}
\end{align}
where $\partial_n$ is the normal derivative with respect to the
worldsheet boundary $\partial \Sigma$.  Therefore,
\begin{align}
 {\delta S_{\text{ren,on-shell}}\over \delta  J}
 &=\sqrt{-\gamma}\biggl(
 -G(\partial_n X+\Box_\gamma X)+\kappa G^2 \left[
 (\nabla X)^2 \partial_n X+3 (\nabla_\gamma X)^2 \Box_\gamma X
 \right]\biggl)
 \notag\\
 &=3x_{(3)}(1+\kappa \dot{J}^2)+\CO(1/r).
\end{align}
In the second equality, we plugged in the explicit expansion
\eqref{FGexpn_X_sol_app}.  Therefore, by the GKPW rule
\cite{Gubser:1998bc, Witten:1998qj}, the expectation value of the
operator $\CO_X=F$ dual to $X$ in the presence of source $x_{(0)}\equiv
J$ is given by, up to $\CO(\kappa^2)$ terms,
\begin{align}
  \ev{F}_J  &= 3x_{(3)}(1+\kappa \dot J^2)
  = 3y_{(3)}+3\kappa \left(z_{(3)}+y_{(3)}\dot J^2\right).
 \label{1ptF_app}
\end{align}
The $\dot J^2$ term may appear strange, but we will see that this term
gets canceled in the final expression for the 4-point function.
Actually, there is a further contribution to \eqref{1ptF_app}, but we
will discuss it later (see below \eqref{Okcomb}).

Although our discussion above was based on the pure AdS space
\eqref{T=0AdS_app} for the simplicity of the argument, the final expression
\eqref{1ptF_app} is valid for general asymptotically AdS space,
including the AdS black hole \eqref{AdSdBH}.  Below, we will use
\eqref{1ptF_app} to compute correlation functions for the AdS black hole
background \eqref{AdSdBH}.

\subsection{Propagators and correlators}
\label{subsec:app:prop_corr}

To compute the expectation value $\ev{F}_J$ using the formula
\eqref{1ptF_app}, we need to know $x_{(3)}=y_{(3)}+\kappa
z_{(3)}+\CO(\kappa^2)$.  This can be determined if we know the
propagators that satisfy appropriate boundary conditions in the inside
of the AdS space as we discuss below.

If we substitute the expansion \eqref{FGexpn_app} into the wave equation
\eqref{EOMgen} and compare the coefficients, we obtain
\begin{subequations}
 \label{eom_obo_app}
 \begin{align}
 [-h^{-1}\partial_t^2 +\partial_r(r^4h\partial_r)]Y&=0,  \label{eom_obo_app0}\\
 [-h^{-1}\partial_t^2 +\partial_r(r^4h\partial_r)]Z&=\rho,  \label{eom_obo_app1}
 \end{align}
\end{subequations}
where we are now considering the AdS black hole spacetime \eqref{AdSdBH}
and the ``source'' $\rho$ is defined by
\begin{align}
 \rho&\equiv -\partial_t (H_0 h^{-1}\partial_t Y)+\partial_r(H_0r^4h\partial_r Y),
 \qquad
 H_0 \equiv -h^{-1}(\partial_t  Y)^2+r^4h(\partial_r Y)^2.\label{def_j,H}
\end{align}
We solve \eqref{eom_obo_app0} under the asymptotic condition $Y(r,t)\to
J(t)$ as $r\to\infty$ and \eqref{eom_obo_app1} under the condition
$Z(r,t)\to 0$ as $r\to \infty$. Let us solve these using propagators.
First, let $K(r,t|t')$ be the boundary-bulk propagator, namely the
solution to the zeroth-order wave equation \eqref{eom_obo_app0}
satisfying the boundary condition
\begin{align}
 K(r,t|t')\to \delta(t-t')\qquad \text{as}\quad r\to\infty.
\end{align}
Then the solution to \eqref{eom_obo_app0} is
\begin{align}
 Y(t,r)=\int dt' K(r,t|t') J(t').
\end{align}
From this, we can read off $y_{(3)}$ as
\begin{align}
 y_{(3)}(t)=\int dt' [K(r,t|t')]|_{r^{-3}}\, J(t').\label{y(3)}
\end{align}
where $[~]_{r^{-3}}$ means to take the coefficient of the $r^{-3}$ term
in the $1/r$ expansion.

Let us move on to the next order equation \eqref{eom_obo_app1} to
determine $z_{(3)}$.  Let $D(r,t|r',t')$ be the bulk propagator, namely
the solution to
\begin{align}
 [-h^{-1}\partial_t^2 +\partial_r(r^4h\partial_r)]D(t,r|t',r')
 =\delta(t-t')\delta(r-r')\label{eom_bulk_prop_app}
\end{align}
that vanishes as $r,r'\to\infty$.  Then the solution to the next order
equation \eqref{eom_obo_app1} can be written as
\begin{align}
 Z(t,r)&=\int dt'\,dr'\, D(t,r|t',r') \rho(t',r').\label{X1=Dj_app}
\end{align}

It is easy to see that the $Z$ given by \eqref{X1=Dj_app} has the
expected behavior \eqref{FGexpn_Z_sol_app}.
To see it, let us explicitly construct the bulk propagator satisfying
\eqref{eom_bulk_prop_app}, or in the frequency space,
\begin{align}
 [h^{-1}\omega^2 +\partial_r(r^4h\partial_r)]D(\omega,r,r') =\delta(r-r').
 \label{eom_bulk_prop_freq_app}
\end{align}
The solution to this can be constructed\footnote{The following argument
is analogous to the one given around \eqref{Dadv_ansatz_app}.} from
the solution to \eqref{eom_obo_app0}, which can be written in the
frequency space as
\begin{align}
  [h^{-1}\omega^2 +\partial_r(r^4h\partial_r)]Y_\omega&=0.  \label{eom_obo_app0freq}
\end{align}
As discussed above the equation \eqref{eom_NH}, this wave equation
\eqref{eom_obo_app0freq} has two solutions; let us denote them by
$\phi_{\pm \omega}(r)$.\footnote{$\phi_{\pm \omega}(r)$ are different
from $g_{\pm\omega}(r)$ defined around \eqref{eom_NH} only by
normalization; $\phi_{\pm \omega}(r)\to 1$ as $r\to \infty$, while
$g_{\pm\omega}(r)\to e^{\pm i\omega r_*}$ as $r\to r_c$
($r_*\to-\infty$). These agree in the small $\omega$ limit.} These are
related to each other by $\phi_{\omega}(r)^*=\phi_{-\omega}(r)$.  As one
can see from \eqref{FGexpn_Y_sol_app}, we can take them to have the
following large $r$ expansion:
\begin{align}
 \phi_{\pm\omega}(r)=1+{\omega^2\over r^2}+{c_{\pm\omega}\over r^3}+\cdots,\label{fomega_inf}
\end{align}
where $c_{\pm \omega}$ are some constants ($c_{\omega}^*=c_{-\omega}$).
For example, in the AdS$_3$ case ($d=3$),
\begin{align}
 \phi_{\pm\omega}(r)=\left(1\pm{i\omega\over r}\right)\left({r-r_H\over
 r+r_H}\right)^{i\omega/2r_H}
 =1+{\omega^2\over 2r^2}\mp {i\omega(r_H^2+\omega^2)\over 3r^3}+\cdots.
\end{align}
%
For $r\neq r'$, the equation \eqref{eom_bulk_prop_freq_app} is the same
as \eqref{eom_obo_app0freq} and therefore $D(\omega,r,r')$ is given by a
linear combination of $\phi_{\omega}(r)$ and $\phi_{-\omega}(r)$.  Taking into
account the $r\leftrightarrow r'$ symmetry, the bulk propagator $D$ can
be written as
\begin{align}
 D(\omega,r,r')&= 
  A\Bigl[ \phi_{\omega}^{>}(r) \phi_{\omega}^{<}(r')\theta(r-r')
 +\phi_{\omega}^{>}(r')\phi_{\omega}^{<}(r)\theta(r'-r)\Bigr].
\label{Dbulk_ansatz_app}
\end{align}
Here $A$ is constant and we defined
\begin{align}
\begin{split}
  \phi_{\omega}^{>}(r)&\equiv \phi_\omega(r)-\phi_{-\omega}(r)={c_\omega-c_{-\omega}\over r^3}+\CO(r^{-4}),\label{asympt_f<>}
 \\
 \phi_{\omega}^{<}(r)&\equiv \alpha \phi_\omega(r)+(1-\alpha)\phi_{-\omega}(r)
 =1+{\omega^2\over 2r^2}+
 {\alpha c_\omega+(1-\alpha )c_{-\omega}\over r^3}+\CO(r^{-4}).
\end{split}
\end{align}
The fact that $\phi_{\omega}^{>}(r)\to 0$ as $r\to 0$ correctly gives the
asymptotic condition for $D$, namely $D\to 0$ as $r,r'\to \infty$.  On
the other hand, we do not specify the boundary condition of $D$ as
$r,r'\to r_H$. The unknown number $\alpha$ parametrizes possible
boundary conditions which is to be determined by some physical
requirement.  But we leave $\alpha$ arbitrary and therefore
\eqref{Dbulk_ansatz_app} is valid regardless of the boundary condition.
Because $\phi_\omega^<(r)\to 1$ as $r\to \infty$, it is actually equal to
the bulk-boundary propagator in the frequency space;
\begin{align}
 \phi_\omega^<(r)=K(\omega,r).\label{f<=K}
\end{align}
By substituting \eqref{Dbulk_ansatz_app} into the equation
\eqref{eom_bulk_prop_freq_app}, we obtain
\begin{align}
 A&={1\over r^4h[\phi_\omega^>(\partial_r \phi_\omega^<)-(\partial_r \phi_\omega^>) \phi_\omega^<]}
\end{align}
(this is the same as \eqref{valueA}).  Since this does not depend on $r$
(see below \eqref{valueA}), by taking $r\to \infty$ and using the
asymptotic behavior \eqref{asympt_f<>}, we find
$A=(c_\omega-c_{-\omega})^{-1}$.  Therefore, the bulk propagator is
found to be
\begin{align}
 D(\omega,r,r')= 
 (c_\omega-c_{-\omega})^{-1}
 \left[
  \phi_{\omega}^{>}(r) \phi_{\omega}^{<}(r')\theta(r-r')
 + \phi_{\omega}^{>}(r')\phi_{\omega}^{<}(r)\theta(r'-r)\right],
\end{align}
where we used \eqref{f<=K}. The $r\to\infty$ behavior of this is, using
the asymptotic behavior \eqref{asympt_f<>},
\begin{align}
 D(\omega,r,r')
 =
 -{1\over 3r^3}K(\omega,r')\theta(r-r')-{1\over 3r'^3}\theta(r'-r)+\CO(r^{-4})
 \qquad (r\to\infty).
 \label{D_asympt_app}
\end{align}
Using \eqref{FGexpn_Y_sol_app}, we can show that the source $\rho$
(defined in Eq.\ \eqref{def_j,H}) goes as
$\rho=2\dot{J}^2\ddot{J}+\CO(r^{-2})$.  Then, from \eqref{X1=Dj_app} and
\eqref{D_asympt_app} we can read off $z_{(3)}$ as follows:
\begin{align}
 z_{(3)}(t)=\lim_{r\to\infty}\left[
 -{1\over 3}\int_{r_H}^r dt'dr' K(t',r'|t)\rho(t',r')
 +{2\over 3}r\dot{J}(t)^2\ddot{J}(t)
 \right].
\label{z(3)}
\end{align}
The second term cancels the divergent contribution corresponding to
$z_{(2)}$ in \eqref{FGexpn_Z_sol_app}.

So, we succeeded in expressing $y_{(3)},z_{(3)}$ appearing in the formula
\eqref{1ptF_app} using propagators; the resulting expressions are
\eqref{y(3)} and \eqref{z(3)}.
Using these, we can compute the boundary correlators for $F$\@.  First,
at the first order in $\kappa$ that we are working in, the 2-point
function gets contribution only from $y_{(3)}$ in \eqref{y(3)} and
\begin{align}
 \ev{\CT[F(t_1)F(t_2)]}
 &=\left.{\delta\over\delta J(t_2)}\ev{F(t_1)}\right|_{J=0}
 =3{\delta\over\delta J(t_2)} y_{(3)}(t_1)
 =3[ K(t_1,r|t_2)]|_{r^{-3}}.
\end{align}
In the frequency space,
\begin{align}
 \ev{F(\omega_1)F(\omega_2)}&=2\pi \delta(\omega_1+\omega_2)\, 3K(\omega_2,r)|_{r^{-3}}.
\end{align}

To obtain 4-point functions, we take functional derivatives of
\eqref{1ptF_app} three times. Therefore, only the second term $3\kappa
(z_{(3)}+y_{(3)}\dot J^2)$ in \eqref{1ptF_app} is relevant for the
computation.  Let us write the source $\rho$ appearing in \eqref{z(3)}
as
\begin{align}
 \rho&=\partial_t \rho^t+\partial_r \rho^r,\qquad
 \rho^t \equiv -H_0 h^{-1}\partial_t Y,\qquad
 \rho^r \equiv H_0 r^4h\partial_r Y.
\end{align}
Then, by partial integration, \eqref{z(3)} becomes
\begin{align}
 z_{(3)}(t)&=\lim_{r\to\infty}\biggl\{
  {1\over 3}\int^r_{r_H} dt'dr' 
 \Bigl[\rho^t(t',r') \partial_{t'} K(t',r'|t)+\rho^r(t',r')
 \partial_{r'} K(t',r'|t)\Bigr]
 \notag\\
 &\qquad\qquad
 -{1\over 3}\int dt' \Bigl[K(t',r|t)\rho^r(t',r)-K(t',r_H|t)\rho^r(t',r_H)\Bigr]
 +{2\over 3}r\dot{J}(t)^2\ddot{J}(t)
 \biggr\}.\label{hbxp26Aug09}
\end{align}
We dropped the boundary terms at $t=\pm\infty$.  The first term in the
second line can be evaluated using the expansion
\begin{align}
 K(t',r|t)=\delta(t-t')+\CO({r^{-2}}),\qquad
 \rho^r(t,r)=r\dot{J}^2\ddot{J}+3y_{(3)}\dot{J}^2+\CO(r^{-1}).
\end{align}
As a result, in the combination appearing in \eqref{1ptF_app}, the term
involving $y_{(3)}\dot{J}^2$ cancels out:
\begin{align}
 3\kappa\left[z_{(3)}(t)+y_{(3)}(t)\dot J(t)^2\right]
 &=\kappa\lim_{r\to\infty}\biggl\{
 \int^r_{r_H} dt'dr' 
 \Bigl[\rho^t(t',r') \partial_{t'} K(t',r'|t)+\rho^r(t',r')
 \partial_{r'} K(t',r'|t)\Bigr]
 \notag\\
 &
  \qquad \qquad \quad 
 +\int dt' K(t',r_H|t)\rho^r(t',r_H)
 +r\partial_t[\dot{J}(t)^3]
 \biggr\}.\label{Okcomb}
\end{align}
The second last term in \eqref{Okcomb} gets canceled by the extra
contribution alluded to below \eqref{1ptF_app}.  Let us now discuss what
this extra contribution is.  The on-shell variation of the action, which
we used to compute the expectation value $\ev{F}_J$, is given by
\eqref{dSren,os}.  Because we are regarding the region $r_H\le r\le r_c$
as our spacetime, there actually is contribution from the ``boundary''
$r=r_H$ to this expression. In the AdS black hole spacetime, this extra
contribution to $\delta S_{\text{ren,on-shell}}$ becomes
\begin{align}
 \delta S_{\text{ren,on-shell}}
 \supset -\int_{r=r_H}dt\,r^4h(\partial_r Y+\kappa H_0\partial_r Y)\delta Y,
\end{align}
where we dropped $\CO(\kappa^2)$ terms and ``$\supset$'' means that the
left hand side includes the expression on the right hand side.  Note
that, because the counter term $S_{\text{ct}}$ \eqref{Sct_app} was added
only for the boundary at infinity, the second and the fourth terms in
\eqref{dSren,os} did not contribute to this expression.  Since $h\to 0$
as $r\to r_H$, this becomes
\begin{align}
 \delta S_{\text{ren,on-shell}}
 &\supset-\kappa \int_{r=r_H}dt\,r^4h H_0\partial_r Y\,\delta Y
\end{align}
(note that $H_0$ involves $h^{-1}$).  Therefore, by taking functional
derivative, we find that there is the following extra contribution to
$\ev{F}_J$:
\begin{align}
 \ev{F(t)}_J=
 {\delta S_{\text{ren,on-shell}}\over \delta J(t)}
 &\supset -\kappa \int_{r'=r_H}dt'\,r'^4h H_0\partial_r Y\,K(t',r'|t).
\end{align}
This precisely cancels the second last term in \eqref{Okcomb}.
Therefore, the terms relevant for computing 4-point functions is
\begin{align}
  {\ev{F(t)}_J}
  &\supset \kappa \lim_{r\to\infty}\biggl\{
 \int^r_{r_H} dt'dr' 
 \Bigl[j_t(t',r') \partial_{t'} K(t',r'|t)+j_r(t',r')
 \partial_{r'} K(t',r'|t)\Bigr]
 +r\partial_t[\dot{J}(t)^3]
 \biggr\}.
 \label{Okcomb2}
\end{align}

By taking functional derivatives of \eqref{Okcomb2}, we find that
\begin{align}
 G^F(t_1,t_2,t_3,t_4)
 &= \ev{\CT[F(t_1)F(t_2)F(t_3)F(t_4)]}
 = \left.{\delta^3\over \delta J(t_2)\delta J(t_3)\delta J(t_4)} \ev{F(t_1)}_J 
 \right|_{J=0}
 \notag\\
 &=\kappa\biggl\{
 {1\over 4} \sum
 \int^{r}_{r_H} dt\,dr
 \left(-{1\over h}\dot K_i \dot K_j + {r^4 h}K'_i K'_j \right)
 \left(-{1\over h}\dot K_k \dot K_l + {r^4 h}K'_k K'_l \right)
 \notag\\
 &\qquad\quad
 +\,6r\, \partial_{t_1}\! \left[\dot\delta(t_1-t_2)\dot\delta(t_1-t_3)\dot\delta(t_1-t_4)\right]
 \biggr\},
\end{align}
where the $r\to \infty$ limit is understood. Also, $K_i\equiv K(t,r|t_i)$
and the summation is over permutations $(ijkl)$ of $(1234)$.  The
expression in the Fourier space is
\begin{multline}
 G^F(\omega_1,\omega_2,\omega_3,\omega_4)
 =2\pi\kappa \delta(\omega_1+\omega_2+\omega_3+\omega_4)\\
 \times\biggl\{
 {1\over 4} \sum_{\text{perm}\atop (ijkl)}
 \int^{r}_{r_H} dr
 \left(\omega_i\omega_j K_i K_j + {r^4 h}K'_i K'_j \right)
 \left(\omega_k\omega_l K_k K_l + {r^4 h}K'_k K'_l \right)
 -6r\omega_1\omega_2\omega_3\omega_4
 \biggr\},\label{4ptfunc_omega_app}
\end{multline}
where now $K_i\equiv K(\omega_i,r)$.  Note that the first term in
\eqref{4ptfunc_omega_app} is the expression for the 4-point function we
would obtain from the naive GKPW rule.  The last term is there to cancel
the UV divergence coming from the first term due to the fact that
$K_i=1+\CO(r^{-2})$.

\subsection{Lorentzian AdS/CFT}
\label{subsec:app:Lorentzian}

So far we have not fully taken into account the fact that our spacetime
is a Lorentzian spacetime, for which we have to use the Lorentzian
AdS/CFT prescription \cite{Skenderis:2008dg, vanRees:2009rw}.

\begin{figure}[tb]
\begin{center}
 \epsfig{file=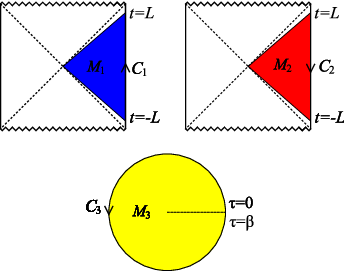,width=7cm} \caption{The bulk geometry
$M=M_1+M_2+M_3$ that ``fills in'' the boundary contour $C=C_1+C_2+C_3$.
For $d>3$, the Penrose diagrams for the Lorentzian patches drawn above
are not accurate because the zigzag singularity lines must actually be
not horizontal but bent inwards \cite{Fidkowski:2003nf}.}
\label{fig:bulkLE}
\end{center}
\end{figure}

On the boundary side, to compute real time correlators, we have to take
the time to run along the contour on the complex place, as we discussed
in subsection \ref{subsec:thermalFT_on_ws}; see Figure
\ref{fig:tcontour} on page \pageref{fig:tcontour}.  The Lorentzian
AdS/CFT prescription is simply to consider a bulk spacetime which
``fills in'' this contour. Then the bulk spacetime will have no boundary
and there is no ambiguity in boundary conditions (although we have to
impose certain gluing condition for fields across different patches).
Following \cite{vanRees:2009rw}, we take the bulk spacetime to be the
union of three patches $M_i$ with $i=1,2,3$, each of which fills in the
corresponding contour $C_i$ in \eqref{C123}.  First, we take $M_1$ to be
the $-L\le t\le L$, $r_H\le r<\infty$ part of the Lorentzian AdS black
hole \eqref{AdSdBH}.  $M_2$ is taken to be the same as $M_1$
metric-wise, but the orientation is taken to be opposite to $M_1$,
corresponding to the fact that $C_1$ and $C_2$ has opposite
orientations.  $M_3$ is taken to be the Euclidean version of the black
hole \eqref{AdSdBH},
\begin{align}
 ds_E^2&={r^2\over l^2}\left[h(r)d\tau^2+(d X^I)^2\right]
 +{l^2\over r^2 h(r)}  dr^2.
\end{align}
The Euclidean time $\tau$ is taken to be $0\le\tau\le \beta$ where
$\beta$ is the inverse Hawking temperature in \eqref{Hawk_temp_d}.  For
a schematic explanation of the patches $M_{1,2,3}$, see Figure
\ref{fig:bulkLE}.
\begin{figure}[tb]
\begin{center}
 \epsfig{file=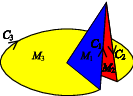,width=4cm} \caption{How to patch together
 the bulk patches $M_1,M_2,M_3$.} \label{fig:patchgeom}
\end{center}
\end{figure}
The way that three patches $M_{1,2,3}$ are glued
together is simply the bulk extension of the way that the contours
$C_{1,2,3}$ are glued together; see Figure \ref{fig:patchgeom}.

Because now our spacetime is not just $M_1$ but $M=M_1+M_2+M_3$, the
action have contributions from all of $M_{1,2,3}$, just as the boundary
\eqref{juii4Oct09}.  Therefore, the bulk integration appearing {\it e.g.}\ in
\eqref{4ptfunc_omega_app} should be now over all $M_i$, with the signs
correctly taken into account:
\begin{align}
 &G^F(\omega_1,\omega_2,\omega_3,\omega_4)
 =2\pi\kappa \delta(\omega_1+\omega_2+\omega_3+\omega_4)
 \notag\\
 &\times\biggl\{
 {1\over 4} \sum_{\text{perm}\atop (ijkl)}
 \int^{r}_{r_H} dr
 \biggl[
 \left({\omega_i\omega_j K_{[11]i} K_{[11]j}\over h} + {r^4 h}K'_{[11]i} K'_{[11]j} \right)
 \left({\omega_k\omega_l K_{[11]k} K_{[11]l}\over h} + {r^4 h}K'_{[11]k} K'_{[11]l} \right)
 \notag\\
 &\qquad\qquad\qquad\quad
 -\left({\omega_i\omega_j K_{[21]i} K_{[21]j} \over h}+ {r^4 h}K'_{[21]i} K'_{[21]j} \right)
  \left({\omega_k\omega_l K_{[21]k} K_{[21]l} \over h}+ {r^4 h}K'_{[21]k} K'_{[21]l} \right)
 \biggr]
 \notag\\
 &\qquad
 -6r\omega_1\omega_2\omega_3\omega_4
 \biggr\}.
 \label{4ptfunc_omega_app2}
\end{align}
Here, $K_{[ab]i}=K_{[ab]}(\omega_i,r)$ and $K_{[ab]}(\omega,r)$ is the
boundary-bulk propagator from the boundary $\partial M_b$ to the bulk
$M_a$.  The second line corresponds to the integration over $M_1$ and
the third line to the integration over $M_2$.  Because we are taking the
$L\to\infty$ limit, the contribution from $M_3$ has been dropped.  The
counter term $-6r\omega_1\omega_2\omega_3\omega_4$ is added only for
$M_1$, because the source is inserted only on $\partial M_1$
($K_{[21]}(\omega,r)$ vanishes as $r\to\infty$).

Because the spacetime $M=M_1+M_2+M_3$ has no boundary inside, the
boundary-bulk propagator can be determined without having to worry about
boundary conditions.  Carefully matching the values across different
patches following \cite{Skenderis:2008dg, vanRees:2009rw}, we find the
boundary-bulk propagators as follows:
 \begin{align}
\begin{split}
  K_{[11]}(\omega,r)&={1\over e^{\beta \omega}-1}[-\phi_\omega(r)+e^{\beta\omega}\phi_{-\omega}(r)],\label{b2bprop_LE_app}\\\
 K_{[21]}(\omega,r)&={e^{\beta \omega}\over e^{\beta \omega}-1}[-\phi_\omega(r)+\phi_{-\omega}(r)],\\
 K_{[31]}(\omega,r)&={e^{(iL+\beta) \omega}\over e^{\beta \omega}-1}[-\phi_\omega(r)+\phi_{-\omega}(r)],
\end{split}
\end{align}
where $\phi_{\pm \omega}(r)$ is the solution to the wave equation
\eqref{eom_obo_app0freq} satisfying the boundary condition
\eqref{fomega_inf}. By substituting these propagators into
\eqref{4ptfunc_omega_app2}, we can finally obtain the 4-point function
for $F$.

\subsection{Low frequency correlators}

We are interested in the low frequency behavior of the correlation
functions.  As we discussed in Appendix \ref{sec:lowE_wave}, the
solution $\phi_{\pm\omega}(r)$ simplifies in the low frequency limit
as\footnote{Note that the precise limit we are taking is
\eqref{low_freq_lim_app}.}
\begin{align}
\phi_{\pm\omega}(r)\sim e^{\pm i \omega r_*}.
\end{align}
If we apply this to \eqref{4ptfunc_omega_app2} and
\eqref{b2bprop_LE_app}, we obtain the following low frequency behavior:
\begin{align}
G^F(\omega_1,\omega_2,\omega_3,\omega_4)
 &\sim {\kappa\over \beta^3}
 \delta(\omega_1+\omega_2+\omega_3+\omega_4)
 \sum\limits_{1\le i<j\le 4}(\omega_i+\omega_j)
 \int_{-\infty}^0 dr_* {r^2\over h}
 e^{-2i (\omega_i+\omega_j)r_*} 
 \notag\\
 &\quad
 +\text{(higher powers in $\omega$)},
\end{align}
where we dropped numerical factors.
Because we have rescaled $X$ in \eqref{Sbare_app-2}, to obtain the
correlator for $F=\CO_X$ dual to the original $X$ before rescaling, we
have to rescale $F\to{ F\over\sqrt{2\pi\ap}}$.  Therefore, in the end,
the 4-point function for $F$ is
\begin{align}
G^F(\omega_1,\omega_2,\omega_3,\omega_4)
 &\sim {1\over \ap\beta^3}
 \delta(\omega_1+\omega_2+\omega_3+\omega_4)
 \sum\limits_{1\le i<j\le 4}(\omega_i+\omega_j)
 \int_{-\infty}^0 dr_* {r^2\over h}
 e^{-2i (\omega_i+\omega_j)r_*} .
\end{align}
This is exactly the same as the result \eqref{R4_expl_naive} that we
obtained by a more naive method.  Namely, this has exactly the same IR
divergence as \eqref{R4_expl_naive} that we studied in section
\ref{sec:IRdiv}.

\subsection{Retarded 4-point function}
\label{apss:retarded4pt}

In the above, we computed the time-ordered 4-point function for the force
$F$ which turned out to be IR divergence. We can also compute the
retarded 4-point function using the above formalism.  As was shown in
\cite{vanRees:2009rw}, for computing retarded correlators, one uses
purely ingoing boundary condition for the boundary-bulk propagator:
\begin{align}
 K_{\text{ret}}(\omega,r)=\phi_{-\omega}(r).
\end{align}
If we define
\begin{align}
 G^F_{\text{ret}}(t_1,t_2,t_3,t_4)
 =\sum_{\text{perm}\atop (ijkl)}\theta(t_i>t_j>t_k>t_l)\ev{[[[F(t_i),F(t_j)],F(t_k)],F(t_l)]}
\end{align}
then the prescription of \cite{vanRees:2009rw} gives
\begin{align}
 &G^F_{\text{ret}}(\omega_1,\omega_2,\omega_3,\omega_4)
 =2\pi\kappa \delta(\omega_1+\omega_2+\omega_3+\omega_4)
 \notag\\
 &\times\biggl\{
 {1\over 4} \sum_{\text{perm}\atop(ijkl)}
 \int^{r}_{r_H} dr
 \left(\omega_i\omega_j K_{\text{ret},i} K_{\text{ret},j} + {r^4 h}K'_{\text{ret},i} K'_{\text{ret},j} \right)
 \left(\omega_k\omega_l K_{\text{ret},k} K_{\text{ret},l} + {r^4 h}K'_{\text{ret},k} K'_{\text{ret},l} \right)
 \notag\\
 &\qquad
 -6r\omega_1\omega_2\omega_3\omega_4
 \biggr\},\label{RRRRret}
\end{align}
where $K_{\text{ret},i}=K_{\text{ret}}(\omega_i,r)$.  The integration
effectively becomes only over $M_1$.

For definiteness, consider the AdS$_3$ case where the retarded correlator is
\begin{align}
 K_{ret}(\omega,r)
 =\left( 1-{i\omega\over r} \right)\left({r-r_H\over r+r_H}\right)^{-i\omega/2r_H}.
\end{align}
For this case, Eq.\ \eqref{RRRRret} gives
\begin{align}
 G^F_{\text{ret}}(\omega_1,\omega_2,\omega_3,\omega_4)
 &=\kappa2\pi\delta(\omega_1+\cdots+\omega_4)\,\omega_1\omega_2\omega_3\omega_4
 \left(2r_H-{16\sum_{i<j}\omega_i\omega_j\over r_H} \right).
\end{align}
Note that this is exact; we have not done low frequency approximation.
This is both IR and UV finite.

\section{Computation of $\eta$ for the STU black hole}
\label{app:eta_STU}

In this Appendix, we will compute the mean-free-path time for the STU
black hole studied in \ref{subsec:app_STU}.  The final results have been
presented in \eqref{tmfp_STU} and \eqref{eta_STU1}--\eqref{eta_STU3}.

We will discuss the 1-charge case
($\kappa_1=\kappa,\kappa_2=\kappa_3=0$) only, because the 2- and
3-charge cases are similar.  First, the relations \eqref{m_rel},
\eqref{T_STU}, and \eqref{Phi_STU} read, in this case,
\begin{align}
 m={r_H^4\over l^2}(1+\kappa),\qquad
 T={r_H\over 2\pi}{2+\kappa\over \sqrt{1+\kappa}},\qquad
 \Phi=-{r_H^2\over \kappa_5^2 l}\sqrt{\kappa}.\label{rels_1chg}
\end{align}

$L_H$ in \eqref{tmfp_STU} can be computed as follows. From the
definition \eqref{L_H_def_text} and \eqref{hthrG_STU} for $n=1$, we obtain
\begin{align}
 \int_\infty^r dr{H^{1/2}\over r^2 f}{1\over \sqrt{1-f^{-1}H^2\CA_t^2}}
&=\int_\infty^r {dr\over r^2-r_H^2}
 \sqrt{r_H^2+\ell^2\over (r^2+r_H^2+\ell^2)((r_H^2+\ell^2)r^2+r_H^4)}.
 \notag\\
&=
 {1\over 2r_H^3}{\sqrt{1+\kappa} \over (2+\kappa)}
 \log{r-r_H\over L_H}+\CO(r-r_H)
\end{align}
The integral in the first line diverges as $r\to r_H$.  We can separate
this divergent piece by subtracting and adding the term obtained by
setting $r$ to $r_H$ in the square root. Further setting $\rho=r/r_H$
and $\kappa=\ell^2/r_H$, we have
\begin{align}
 &{1\over 2r_H^3}{\sqrt{1+\kappa} \over (2+\kappa)}
 \left\{
 \log{r-r_H\over r+r_H}
+ 2(2+\kappa)\int_\infty^1 {d\rho\over \rho^2-1}\left[
 {1\over \sqrt{(\rho+1+\kappa)((1+\kappa)\rho^2+1)}}-
 {1\over (2+\kappa)}\right]
 \right\}\notag\\
&
 \equiv
 {1\over 2r_H^3}{\sqrt{1+\kappa} \over (2+\kappa)}
 \log{r-r_H\over L_H}+\CO(r-r_H).
\end{align}
In the second term in the first line, we have set the upper limit of the
integral to $\rho\to 1$ (which is equivalent to $r\to r_H$) because the
integral is now convergent.  By comparing both sides, we obtain
\begin{align}
 L_H=2r_H\exp\left\{
 -2(2+\kappa)\int_\infty^1 {d\rho\over \rho^2-1}\left[
 {1\over \sqrt{(\rho^2+1+\kappa)((1+\kappa)\rho^2+1)}}-
 {1\over 2+\kappa}\right]
 \right\}.
\end{align}
By using \eqref{rels_1chg}, we obtain the final expression
\eqref{eta_STU1}.  For small $\kappa$, it is easy to expand the
integrand in \eqref{eta_STU1} in $\kappa$, and each integral converges.
This leads to the following expansion of $\eta$ in $\kappa$:
\begin{align}
\eta=
 e^{-\pi/2}\left[
 2 \pi - \pi  \kappa +\frac{(12-\pi) \pi }{16}   \kappa ^2\right]
 +\CO\left(\kappa ^3\right)
\label{eta_STU_expn1}
\end{align}
This shows that, as $\kappa$ increases with fixed $T$, the
mean-free-path time $t_{mfp}$ increases.

The 2- and 3-charge cases are similar and we obtain \eqref{eta_STU2} and
\eqref{eta_STU3}.  The small $\kappa$ expansion of $\eta$ is
\begin{align}
\label{eta_STU_expn2}
\eta&=
e^{-\pi /2}\left[2  \pi -\frac{1}{2} (4-\pi) \pi  \kappa +\frac{1}{16} \pi  \left(52-19 \pi +\pi ^2\right) \kappa
   ^2+O\left(\kappa ^3\right)\right]\\
\label{eta_STU_expn3}
\eta&=
e^{-\pi /2}\left[2  \pi  +(\pi-3) \pi  \kappa +\frac{1}{16} \pi  \left(140-57 \pi +4 \pi ^2\right) \kappa ^2+O\left(\kappa ^3\right)\right]
\end{align}
for the 2- and 3-charge cases, respectively.


\end{document}